\documentclass[11pt,preprint]{aastex}
\usepackage{amsmath}
\usepackage{graphicx}
\usepackage{ulem}
\usepackage{natbib}
\usepackage{lineno}

\usepackage{xspace}
\usepackage{color}

\newcommand{\g}{$\gamma$\xspace}
\newcommand{\hi}{H$\,${\sc i}\xspace}
\newcommand{\hii}{H$\,${\sc ii}\xspace}

\newcommand{\nhi}{$N_{\text{H}\,\textsc{i}}$\xspace}

\newcommand{\fb}{\textit{Fermi} bubbles\xspace}
\newcommand{\loopI}{Loop~I\xspace}
\newcommand{\xco}{$X_{\text{CO}}$\xspace}

\definecolor{seagreen}{rgb}{0.125, 0.70, 0.67}

\definecolor{red}{rgb}{0.9, 0., 0.}

\begin{document}
%\setpagewiselinenumbers
\modulolinenumbers[1]
%%%%%%%%%%%%%%\linenumbers

\title{Development of the Model of Galactic Interstellar Emission for Standard Point-Source Analysis of {\it Fermi} Large Area Telescope Data}

%%%%%%%%%%%%%%%%%%%%%%%%
\author{
F.~Acero\altaffilmark{1}, 
M.~Ackermann\altaffilmark{2}, 
M.~Ajello\altaffilmark{3}, 
A.~Albert\altaffilmark{4}, 
L.~Baldini\altaffilmark{5,4}, 
J.~Ballet\altaffilmark{1}, 
G.~Barbiellini\altaffilmark{6,7}, 
D.~Bastieri\altaffilmark{8,9}, 
R.~Bellazzini\altaffilmark{10}, 
E.~Bissaldi\altaffilmark{11}
% ...
}
\altaffiltext{1}{Laboratoire AIM, CEA-IRFU/CNRS/Universit\'e Paris Diderot, Service d'Astrophysique, CEA Saclay, F-91191 Gif sur Yvette, France}
\altaffiltext{2}{Deutsches Elektronen Synchrotron DESY, D-15738 Zeuthen, Germany}
\altaffiltext{3}{Department of Physics and Astronomy, Clemson University, Kinard Lab of Physics, Clemson, SC 29634-0978, USA}
\altaffiltext{4}{W. W. Hansen Experimental Physics Laboratory, Kavli Institute for Particle Astrophysics and Cosmology, Department of Physics and SLAC National Accelerator Laboratory, Stanford University, Stanford, CA 94305, USA}
\altaffiltext{5}{Universit\`a di Pisa and Istituto Nazionale di Fisica Nucleare, Sezione di Pisa I-56127 Pisa, Italy}
\altaffiltext{6}{Istituto Nazionale di Fisica Nucleare, Sezione di Trieste, I-34127 Trieste, Italy}
\altaffiltext{7}{Dipartimento di Fisica, Universit\`a di Trieste, I-34127 Trieste, Italy}
\altaffiltext{8}{Istituto Nazionale di Fisica Nucleare, Sezione di Padova, I-35131 Padova, Italy}
\altaffiltext{9}{Dipartimento di Fisica e Astronomia ``G. Galilei'', Universit\`a di Padova, I-35131 Padova, Italy}
\altaffiltext{10}{Istituto Nazionale di Fisica Nucleare, Sezione di Pisa, I-56127 Pisa, Italy}
\altaffiltext{11}{Istituto Nazionale di Fisica Nucleare, Sezione di Bari, I-70126 Bari, Italy}
% ...
%%%%%%%%%%%%%%%%%%%%%%%%

%%%%%%%%%%%%%%   %\date{\today\\Version 20}

\begin{abstract}
Most of the celestial $\gamma$ rays detected by the Large Area Telescope (LAT) aboard the {\it Fermi Gamma-ray Space Telescope} originate from the interstellar medium when energetic cosmic rays interact with interstellar nucleons and photons. Conventional point and extended source studies rely on the modeling of this diffuse emission for accurate characterization. 
%Studying $\gamma$ rays from interstellar regions is also a powerful indirect tool to probe cosmic rays, the interstellar gas and the radiation field from where they originate. 
We describe here the development of the Galactic Interstellar Emission Model (GIEM) that is the standard adopted by the LAT Collaboration and is publicly available. The model is based on a linear combination of maps for interstellar gas column density in Galactocentric annuli and for the inverse Compton emission produced in the Galaxy.
%We used all-sky radio and millimeter-wave surveys to trace the hydrogen column density that we corrected with a dust tracer and used an inverse Compton spatial distribution prediction from the propagation code GALPROP. 
%To allow for a Galactocentric gradient of cosmic-ray density in the Galaxy we partitioned the gas column density maps into Galactocentric annuli. The spectral energy distributions (SEDs) of the $\gamma$-ray emission associated with each emission component are determined from fits to 4 years of LAT data in 14 independent energy bins from 50 MeV to 50 GeV. We fit those SEDs with a realistic model for the radiation processes and extrapolate to higher energies.
We also include in the GIEM large-scale structures like \loopI and the \fb. 
%In the absence of suitable templates at other wavelengths, we follow an iterative procedure based on the re-injection of LAT residual maps above the best-fit expectation, filtered to include angular structures on scales larger than 2$\degr$. 
The measured gas emissivity spectra confirm that the cosmic-ray proton density decreases with Galactocentric distance beyond 5~kpc from the Galactic Center. The measurements also suggest a softening of the proton spectrum with Galactocentric distance.
We observe that the 
%North and South 
\fb have boundaries with a shape similar to a catenary at latitudes below 20$\degr$ and we observe an enhanced emission toward their base extending in the North and South Galactic direction and located within $\sim$4$\degr$ of the Galactic Center. 
%We also observe a soft $\gamma$-ray emission with broad extents in the first and fourth quadrant from unknown origin.  
%An excess of $\gamma$-ray , compared to the intensity in the bubbles at higher latitude, is apparent at their bases. 
%Apart from large scale structure like \loopI and the {\it Fermi} bubbles, we observed an broad $\gamma$-ray emission distributed along the plane at longitude less than 50$\degr$ and to a less extent at longitudes around 315$\degr$, this emission is not correlated with any of the templates derived from observations at other wavelengths.

\end{abstract}
\maketitle

\section{Introduction}
\label{sec:Introduction_section}
The hypothesis of interstellar $\gamma$-ray emission, also known as diffuse Galactic emission, dates back to the 1950s when Satio Hayakawa suggested the existence of intense photon production resulting from the decay of the newly discovered neutral pion \citep{Hayakawa:1952p3259}.  Early estimates of the intensity and distribution of this emission, together with the bremsstrahlung radiation of electrons and positrons, inverse-Compton scattering (IC), as well as other secondary mechanisms responsible for the production of interstellar emission, were made by \cite{Morrison:1958p4134}, \cite{Pollack:1963p4127}, \cite{Ginzburg:1965p3189}, and \cite{Stecker:1966p2619}. 

The Third Orbiting Solar Observatory {\it OSO-3}, launched in 1967, confirmed the existence of an Galactic interstellar emission by observing for the first time a correlation between high-energy $\gamma$ rays and Galactic structures \citep{Kraushaar:1972p3349}. Then the Small Astronomy Satellite 2 ({\it SAS-2}, launched in 1972), collecting 20 times more photons, provided clear evidence of correlation between the distributions of high-energy $\gamma$ rays and of atomic hydrogen (\hi). This evidence was quantified by \cite{Lebrun:1979p3085} who compared the {\it SAS-2} sky intensity with the atomic hydrogen column density ($N_{\text{H}\textsc{~i}}$). Later, the comparison between the Cosmic Ray Satellite {\it COS-B} data and the dust extinction derived from galaxy counts revealed the contribution of the molecular hydrogen gas to the  interstellar emission \citep{Lebrun:1982p3092, Strong:1982p4071}. When the wide-latitude Columbia University CO survey \citep{Dame:1984p4026} became available, \cite{Lebrun:1983p3810} used it as a tracer for molecular hydrogen. In \cite{Bloemen:1986p3811} and \cite{Strong:1988p3093} the high-energy interstellar emission was realistically modeled for the first time for the whole Galaxy with the inclusion of IC and a partitioning of the gas column density into four Galactocentric annuli to account for radial variations in cosmic-ray (CR) density. 

The work described in the present paper is in part based on a similar template method. In this approach we do not calculate the intensity of the model components from assumed CR densities and production cross-sections. We use instead the spatial correlation between the $\gamma$-ray data and a linear combination of gas and IC maps in order to (i) model the diffuse background for point-source studies, and (ii) estimate the $\gamma$-ray emissivity of the gas in different regions across the Galaxy.
%a linear combination of templates, which are maps spatially correlated with production sites of $\gamma$ rays. 
The intensity associated with each template is determined from a fit to $\gamma$-ray data. \cite{Strong:1996p1032} successfully applied this method to observations of the Energetic Gamma Ray Experiment Telescope ({\it EGRET}) on the {\it Compton Gamma-Ray Observatory}. Launched in April 1991, {\it EGRET} provided a higher angular and energy resolution and collected about four times more  $\gamma$ rays than {\it COS-B} above 100~MeV. This method was extended by \cite{Casandjian:2008p1563} to include the dark neutral medium (DNM) gas \citep{Grenier:2005p836, Ade:2011p3178, Ade:2014p4235} and to study the influence of the interstellar emission model on the detection of $\gamma$-ray sources. 

The Large Area Telescope (LAT)  \citep{Atwood:2009p4055} is the main $\gamma$-ray detector of the {\it Fermi Gamma-ray Space Telescope} ({\it Fermi}) launched on 2008 June 11. Its pair-production towers collect $\gamma$ rays in the energy range of 20~MeV to greater than 300~GeV. {\it Fermi} was operated in all-sky survey mode for most of its first 4 years of operation, allowing the LAT, with its wide field of view of about 2.4~sr, to image the entire sky every two orbits (or three hours). The survey mode, together with an on-axis effective area of $\sim$8000~cm$^2$ and a 68\% containment of the point-spread function (PSF) of 0\fdg8 at 1~GeV, make the LAT data well suited for studies of interstellar emission and large-scale structures. 
%an interstellar emission model based on the observations. 

\begin{figure}
\begin{center}
\includegraphics[width=16cm]{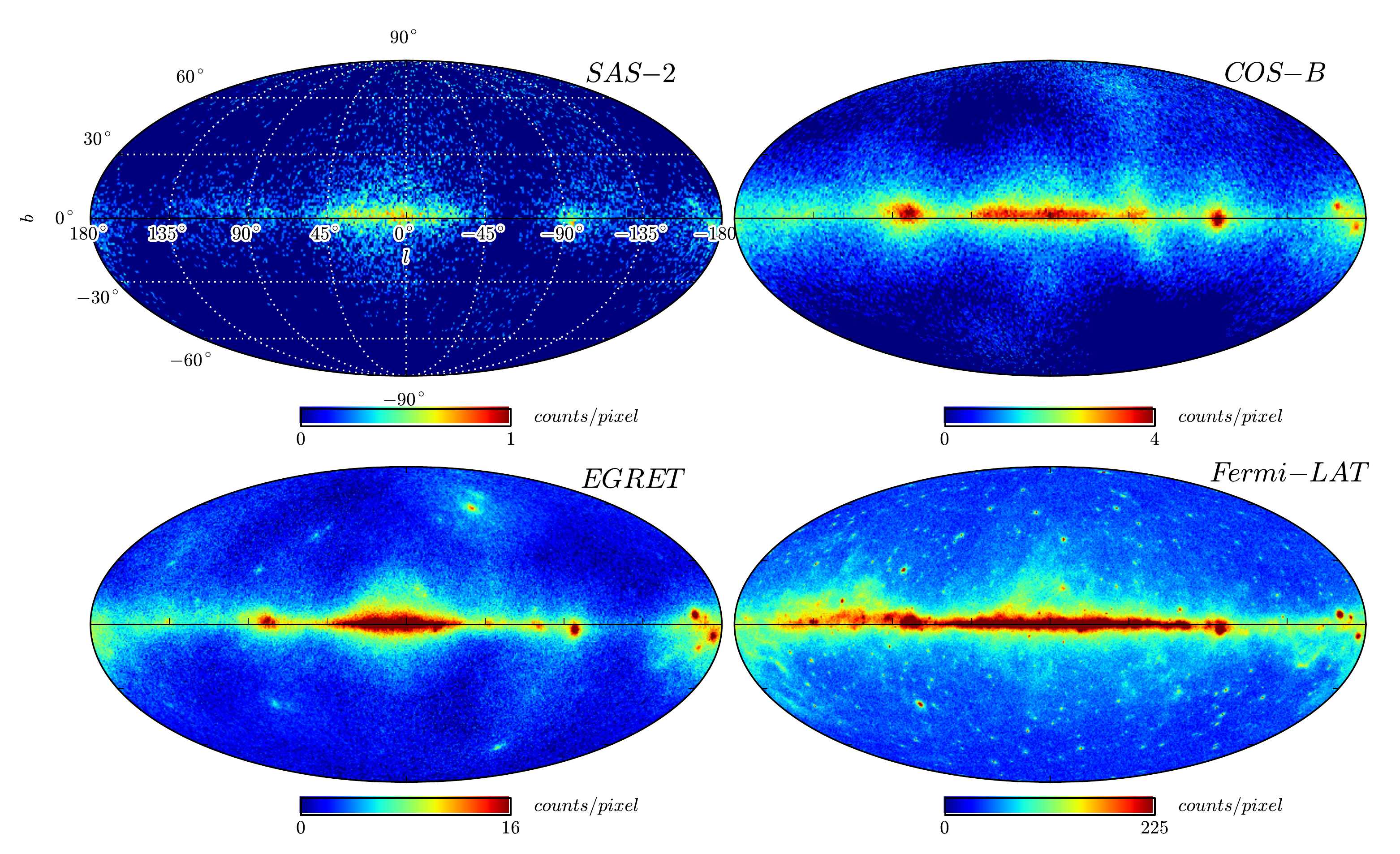}
\caption{Mollweide projection in Galactic coordinates of accumulated counts maps for {\it SAS-2}, {\it COS-B}, {\it EGRET} (above 50~MeV) and {\it Fermi}-LAT (above 360~MeV, 4 years, Clean class events). 
%The ratios of accumulated counts above 50~MeV for successive experiments are : SAS-2/OSO-3=19, COS-B/SAS-2=17, EGRET/COS-B=6, {\it Fermi}-LAT/EGRET=37. 
Regions with enhanced numbers of counts due to a non-uniform exposure time in observations with pointed observations are apparent in panels corresponding to {\it SAS-2}, {\it COS-B}, and {\it EGRET}.}
\label{history}
\end{center}
\end{figure}

Figure \ref{history} shows 4 years of LAT data together with observations\footnote{\url ftp://heasarc.gsfc.nasa.gov} from {\it SAS-2}, {\it COS-B}, and {\it EGRET}. At LAT energies, the diffuse $\gamma$-ray emission of the Milky Way dominates the sky. It contributes five times more photons above 50~MeV than point sources, half of them originating from within 6$\degr$ of the Galactic midplane. The diffuse emission is bright and structured, especially at low Galactic latitudes, and is a celestial back/foreground for detecting and characterizing $\gamma$-ray point sources. Standard LAT analyses based on model-fitting techniques to study discrete sources of $\gamma$ rays require an accurate spatial and spectral model for the Galactic diffuse emission. The LAT Collaboration has previously released two Galactic Interstellar Emission Model (GIEM) versions based on the template approach corresponding to the {\it gll\_iem\_v02.fit}\footnote{\url{http://fermi.gsfc.nasa.gov/ssc/data/access/lat/ring\_for\_FSSC\_final4.pdf}} and  {\it gal\_2yearp7v6\_v0.fits}\footnote{\url{http://fermi.gsfc.nasa.gov/ssc/data/access/lat/Model\_details/Pass7\_galactic.html}} tuned respectively to 10 months and 24 months of observations. The template method was also applied to LAT observations for studying the interstellar emission in several dedicated regions \citep{Abdo:2010p3307, Ackermann:2011p4125, Ackermann:2012p4116, Ackermann:2012p4119, Ade:2014p4235, Ade:2014p4235}. This paper describes the GIEM recommended for point source analyses of the LAT Pass 7 reprocessed data (P7REP) where events have been reconstructed using updated calibrations for the subsystems of the LAT \citep{Bregeon:2013p4093}. 

An alternative method to model the interstellar emission consists in a priori calculations of the CR density and folding this density with $\gamma$-ray production cross-sections. This method was used to derive the official {\it EGRET} interstellar model \citep{Bertsch:1993p3161}. The model was based on the assumption that the CR density distribution follows the density of matter convolved with a two-dimensional Gaussian distribution whose width, representing the matter and CR coupling scale, was left as an adjustable parameter in a fitting procedure applied to {\it EGRET} observations. In this model the molecular-hydrogen-to-CO conversion factor (\xco) was also left free to vary. \cite{Hunter:1997p329} found a good agreement between the model and  {\it EGRET} observations except for an excess of $\gamma$ rays observed above 1~GeV. This excess is likely of instrumental origin \citep{Stecker:2008p4103} and is not seen with the LAT \citep{Abdo:2009p4020}. 

The coupling between CRs and matter assumed in \cite{Bertsch:1993p3161} may not capture the details of CR propagation in the Galaxy. It predicted a large contrast in $\gamma$-ray emissivity in and outside of spiral arms, at odds with the observation \citep{Digel:2001p342,Ackermann:2011p4125}. An alternative approach to estimate the CR density uses a CR propagation code like GALPROP\footnote{\url{http://galprop.stanford.edu}} \citep{Strong:2007p1625} to model the distribution of CRs across the Galaxy. This code was used in  \cite{Strong:2000p306} and \cite{Strong:2004p336}  to predict the Galactic interstellar $\gamma$-ray intensity based on the local CR measurements and assumed distribution of CR sources. GALPROP is widely used by the high-energy astrophysics community; recent advances are described in \cite{Vladimirov:2011p4099}. An extensive comparison between the interstellar emission detected by LAT and GALPROP predictions was performed by \cite{Ackermann:2012p2978}. The authors ran GALPROP with various input parameter sets sampled within realistic ranges. They obtained a grid of models associated with different CR source tracers and with various radial and vertical Galactic boundaries. Unlike the present work, the models were not tuned on LAT data except for \xco. They found that the GALPROP models were broadly consistent with LAT data, but noted an under-prediction of the diffuse $\gamma$-ray emission in the inner Galaxy above a few GeV and a need for higher CR flux or gas density in the outer Galaxy \citep{Abdo:2010p3307}.
%This latter point was interpreted in terms of an inhomogeneous CR diffusion coefficient by \cite{Evoli:2012p4056} and \cite{Gaggero:2013p4062} using the DRAGON\footnote{\url{http://www.desy.de/~cevoli/DRAGON/Home.html}} CR propagation code. Alternatively, since the steady state approximation holds for CR particles, one can calculate the CR density with the simple leaky box model \citep{Maurin:2001p3170}. In this approach the particles freely propagate inside a volume until they lose their energy or escape. \cite{Delahaye:2011p2770} used this method to derive proton densities that they folded with $\gamma$-ray production cross-section and compared to LAT observations. 

Reasonable agreements have been obtained between propagation code predictions and observations at various wavelengths \citep{Strong:2007p1625, Strong:2011p2672, Bouchet:2011p3355, Orlando:2013p4195, Gaggero:2013p4397}, but our knowledge of the distribution of CR sources, of injection spectra, of CR diffusion properties, and of $\gamma$-ray production cross-sections is not accurate enough for these models to be used to precisely describe the diffuse background for point- and extended-source analysis. This is illustrated for example in Figure 6 of \cite{Ackermann:2012p2978}. The $\gamma$-ray emissivity of the gas, which provides the dominant component of the interstellar emission at the energies considered here, is proportional to the CR density folded with $\gamma$-ray production cross sections. The emissivities are more accurately determined with a fit to the data using the template method than from a priori predictions. Nevertheless propagation codes like GALPROP are essential for the calculation of the spatial distribution of the IC for which no template exists. This emission component is present in every direction of the sky but is brightest in the inner Galactic plane.
It is present in the $\gamma$-ray data at large angular scales and it dominates the diffuse emission at energies below ${\sim}\,100$~MeV. It is therefore important to include a prediction for its spatial distribution in order to properly fit other large-scale components of the GIEM, such as the atomic gas.

The interstellar emission can also be studied directly from observations without using spatial templates. \cite{Selig:2015p4426} partitioned the Fermi-LAT counts map above 0.6 GeV into point-like and diffuse contributions. They observed a soft diffuse component tracing the gas content of the interstellar medium (ISM) and a harder one interpreted as IC.

In Section \ref{sec:Model_description_section} we detail the basic ingredients of the GIEM and the derivation of its templates from observations at other wavelengths. The model has free parameters that are fitted to a selection of $\gamma$ rays described in Section \ref{sec:Data_Selection_section}. 
The model equation and parameters are given in Section  \ref{sec:fitting_procedure_section}. 
The fitting procedure itself and the interpretation of the derived emissivities are detailed in Section \ref{sec:fit_and_interpretation}. 
The IC intensities predicted with GALPROP for Galactic electrons and positrons need spectral modification as explained in Section \ref{sec:ICnorm}.
%In Section \ref{sec:High_energy_extrapolation_section} we explain how we \isa{have} extrapolated the fit parameters to higher energies. 
We show how we have modeled the part of the diffuse emission that does not correlate with any of the templates in Section \ref{extra_emission}. We finally describe the construction of the overall GIEM and compare it to LAT observations in Section \ref{sec:Resulting_model_section}.

\section{Emission components}
\label{sec:Model_description_section}
The high-energy interstellar $\gamma$-ray emission is produced by the interaction of energetic CRs with interstellar nucleons and photons. The decay of secondary particles produced in hadron collisions, the IC of the interstellar radiation field (ISRF) by electrons, and their bremsstrahlung emission in the interstellar gas are the main contributors to the diffuse Galactic emission. Underlying our modeling efforts is the reasonable assumption that energetic CRs uniformly penetrate all gas phases in the ISM. 
CR transport and interactions in magnetized molecular clouds involve complex focusing and magnetic mirror effects and diffusion on small-scale magnetic fluctuations that may lead to an exclusion of CRs from the clouds, or conversely to their concentration in the clouds \citep{Skilling:1976p4058,Cesarsky:1978p4059,Gabici:2007p4060,Everett:2011p4057,Padovani:2013p4061}. These processes modify the CR flux at low energies relevant for gas ionization, but they should leave the CR flux at energies above a few GeV unchanged. We thus do not expect changes in $\gamma$-ray emissivity through the rather diffuse gas phases which hold most of the mass, i.e. the HI-bright to CO-bright phases. Experimentally, no evidence of CR screening or re-acceleration in molecular clouds was observed in molecular complexes or local regions studied with the template method \citep{Digel:1999p335,Digel:2001p342,Abdo:2010p3307,Ade:2014p4235}, except for the Cygnus region where a `cocoon' of freshly accelerated particles was observed with a limited spatial extension of approximately 2$\degr$ \citep{Ackermann:2011p2959}. We have therefore assumed that the diffuse $\gamma$-ray intensity at any energy can be modeled as a linear combination of templates of hydrogen column density, assuming a uniform CR distribution within each one, an IC intensity map predicted by GALPROP ($I_{IC_{p}}$), a template that partially accounts for the emission from \loopI ($I_{LoopI}$), an isotropic intensity that accounts for unresolved extragalactic $\gamma$-ray sources and for residual CR contamination in the photon data, a map of the solar and lunar emissions and one for the Earth's limb emission reconstructed in the tails of the LAT point-spread function. In order to compare with the all-sky survey data, the model also includes point-like and extended sources from a preliminary version of the 3FGL catalog \citep{Ballet:2013p4107}.

\subsection{Gas column densities}

About 99\% of the ISM mass is gas and about 70\% of this mass is hydrogen. The hydrogen gas exists in the form of neutral atoms in cold and warm phases, in the form of neutral H$_2$ molecules, and in an ionized state (diffuse H$^{+}$ and \hii regions) \citep{Heiles:2003p1829}. Helium and heavier elements are considered to be uniformly mixed with the hydrogen. 
The warm \hi medium and, to some extent, the cold \hi medium are traced by 21-cm line radiation. Most of the cold molecular mass is traced by $^{12}$CO line emission. At the interface between the atomic and molecular phases, a mixture of dense \hi and diffuse H$_2$ escapes the \hi and CO radio surveys because this intermediate medium is optically thick to \hi photons and CO molecules are absent or weakly excited. This dark neutral medium (DNM) can be indirectly traced by its dust and CR content \citep{Grenier:2005p836,Ade:2014p4235}.

\subsubsection{Atomic hydrogen}
\label{sec:Atomic_hydrogen}
We have derived the atomic column density (\nhi) maps from the 21-cm line radiation temperature under the assumption of a uniform spin (excitation) temperature ($T_{S}$). We used the 21-cm all-sky Leiden-Argentine-Bonn (LAB) composite survey of Galactic \hi \citep{Kalberla:2005p3048} to determine the all-sky distribution of \nhi. The LAB survey merges the Leiden/Dwingeloo Survey (LDS) of the sky north of $\delta = -30\degr$ with the Instituto Argentino de Radioastronomia Survey (IAR) of the sky south of $\delta= -25\degr$. The spatial resolution after regridding is  35$\arcmin$ in the case of the IAR survey, and  40$\arcmin$ in the case of the LDS. We have derived \nhi from the observed brightness temperature $T_B$ with Equation \ref{eqNH}:
\begin{equation}
\begin{split}
%N_{\text{H}\,\textsc{i}}=1.82\times10^{18}~T_S\int_v{ln (1-T_{B}(v)/(T_{S}-2.66))^{-1}} dv   ~~[\textrm{cm}^{-2}]
N_{\text{H}\,\textsc{i}}=-1.82\times10^{18}~T_S\int_v{\ln \left[ 1-\frac{T_{B}(v)}{T_{S}-T_0} \right]} \, dv   ~~[\textrm{cm}^{-2}]
\end{split}
\label{eqNH}
\end{equation}
where the integral is taken over the velocity range of interest \citep{Ackermann:2012p2978} and $T_0 = 2.66$~K represents the background brightness temperature in this frequency range. 

\begin{figure}
\begin{center}
\includegraphics[width=12cm]{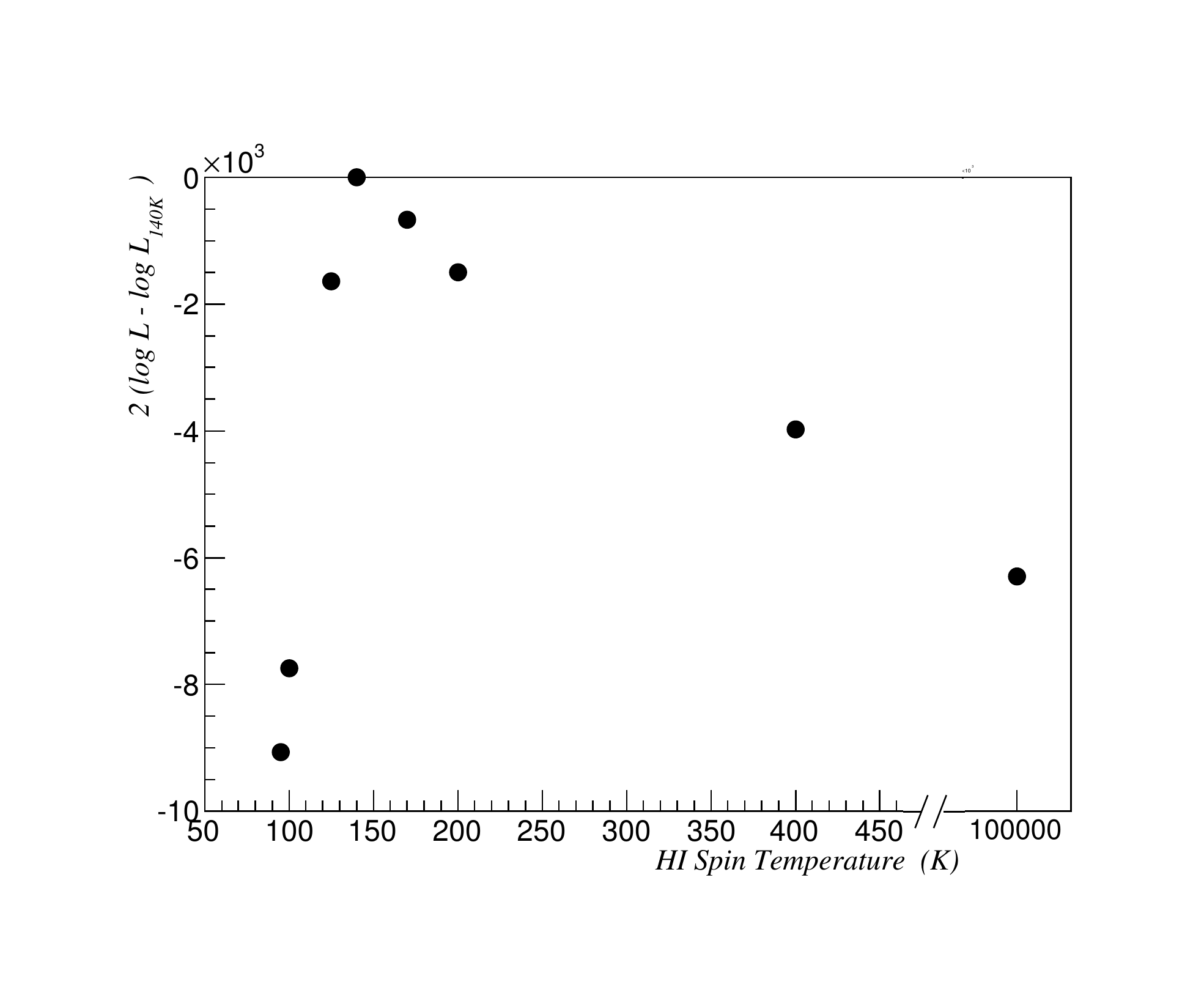}
\caption{Evolution, with the \hi spin temperature $T_S$, of the log-likelihood obtained for the best fits of the interstellar model to the LAT data in the anticenter region ($90\degr \le l \le 270\degr$ and $\left| b \right| <70\degr$) at all energy ranges. The log-likelihood ratio is given with respect to the very best fit obtained at $T_S = 140$~K. The temperature $T_{S}=10^5$~K is equivalent to the optically thin approximation.
%Log-likelihood ratio corresponding to the best fit of the interstellar model (Equation \ref{eqRing}) to {\it Fermi}-LAT data versus the atomic hydrogen spin temperature $T_{S}$. We incorporated in the model a $N_{\text{H}\textsc{~i}}$ template derived from the 21-cm LAB survey assuming  various $T_{S}$. We restricted the fit to the region $90\degr \le l \le 270\degr$ and $\left| b \right| <70\degr$. We set the reference of the log-likelihood to its maximum value observed for $T_{S}=140$~K. The temperature $T_{S}=10^5$~K is equivalent to the optically thin approximation. 
}
\label{TS}
\end{center}
\end{figure}

Since the cold \hi clumps are embedded in the more diffuse warm gas, emission from both cold and warm atomic media can be detected in the same line of sight. Studies of \hi absorption against background radio sources have shown that $T_{S}$ is not uniform in the multi-phase ISM \citep{Heiles:2003p1829, Kanekar:2011p4066}. \cite{Heiles:2003p1829} found that most of the \hi in the cold neutral medium has a $T_{S}$ of less than 100~K and that the warm neutral medium gas lies roughly equally in the thermally unstable region (500$-$5000~K) and in the stable phase above 5000~K. In the absence of $T_S$ information outside the small samples of background radio sources we cannot predict the variations of $T_S$ across the \hi LAB survey. Instead, we have selected the single uniform temperature that provided the best fit to the LAT data (see Section \ref{sec:Data_Selection_section}) in the anticenter region, at $90\degr \le l \le 270\degr$ and $\left| b \right| <70\degr$. Changes in $T_S$ indeed modify the spatial distribution of \nhi as seen from the Earth and these changes can be probed by the \g-ray emission produced by CRs in the atomic gas. The anticenter region was chosen because the uncertain IC emission is dimmer in the outer Galaxy, because these directions are free of large and bright extended sources unrelated to the gas, such as \loopI and the \fb, and because the Galactic warp creates a characteristic signature in the gas maps beyond the solar circle. In that region the LAT observations of the diffuse emission are accurately reproduced from the gas, dust and IC distributions. As we describe below, we have traced additional hydrogen in the DNM from dust column densities. For a uniform dust-to-gas ratio and uniform grain radiative emissivity, the DNM map partially corrects for local decreases in $T_S$. We have investigated seven values of $T_{S}$ from 90~K to 400~K. For each temperature, we have derived \hi and DNM column-density maps (see Section \ref{sec:Dark_neutral_medium}). Figure \ref{TS} shows the log-likelihood ratio obtained by fitting the seven models at all \g-ray energies.
%We observed a maximum in the likelihood plot, 
The model obtained with $T_{S}=140$~K gives the best fit to the LAT photon maps. This average temperature is smaller than the 250-400~K values found with \hi absorption and emission spectra at 1$\arcmin$--2$\arcmin$ resolution beyond the solar circle \citep{Dickey:2009p1250}, but it falls well within the temperature distribution found in the inner Galaxy, which peaks between 100~K and 200~K \citep{Dickey:2003p801}, so we have adopted this temperature of 140~K to derive \hi column densities in the whole Galaxy in order to model the LAT data. We postpone dedicated studies of \hi complexes in the outer Galaxy to understand why we find a larger fraction of cold \hi than the radio studies, even though our model includes additional DNM gas.
%We do not claim that $T_{S}$=140~K corresponds to an average spin temperature in the outer Galaxy. 

\begin{figure}
  \centering
  \includegraphics[width=0.48\textwidth]{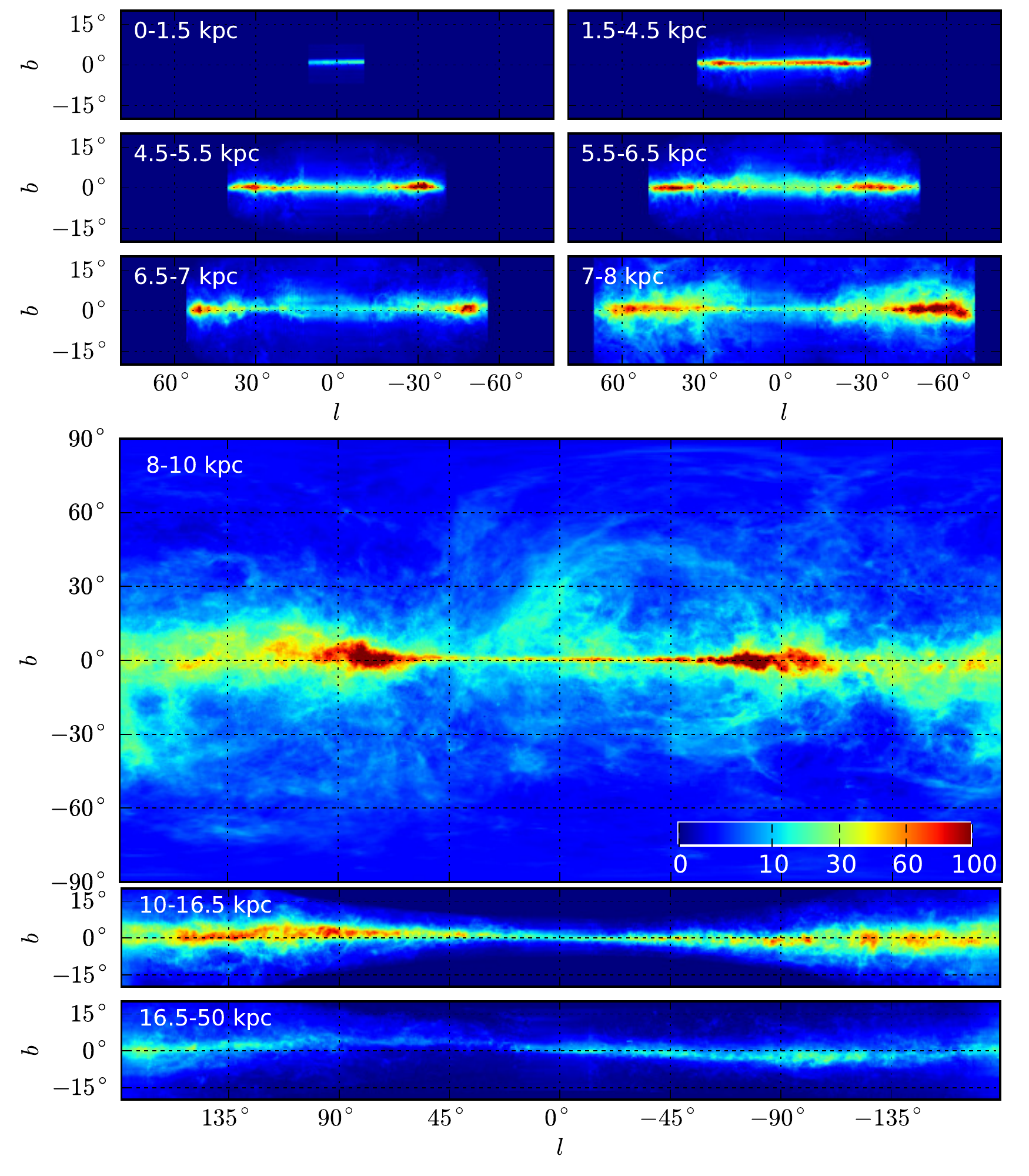}
%  \caption{$N_{\text{H}\textsc{~i}}$}
  \includegraphics[width=0.48\textwidth]{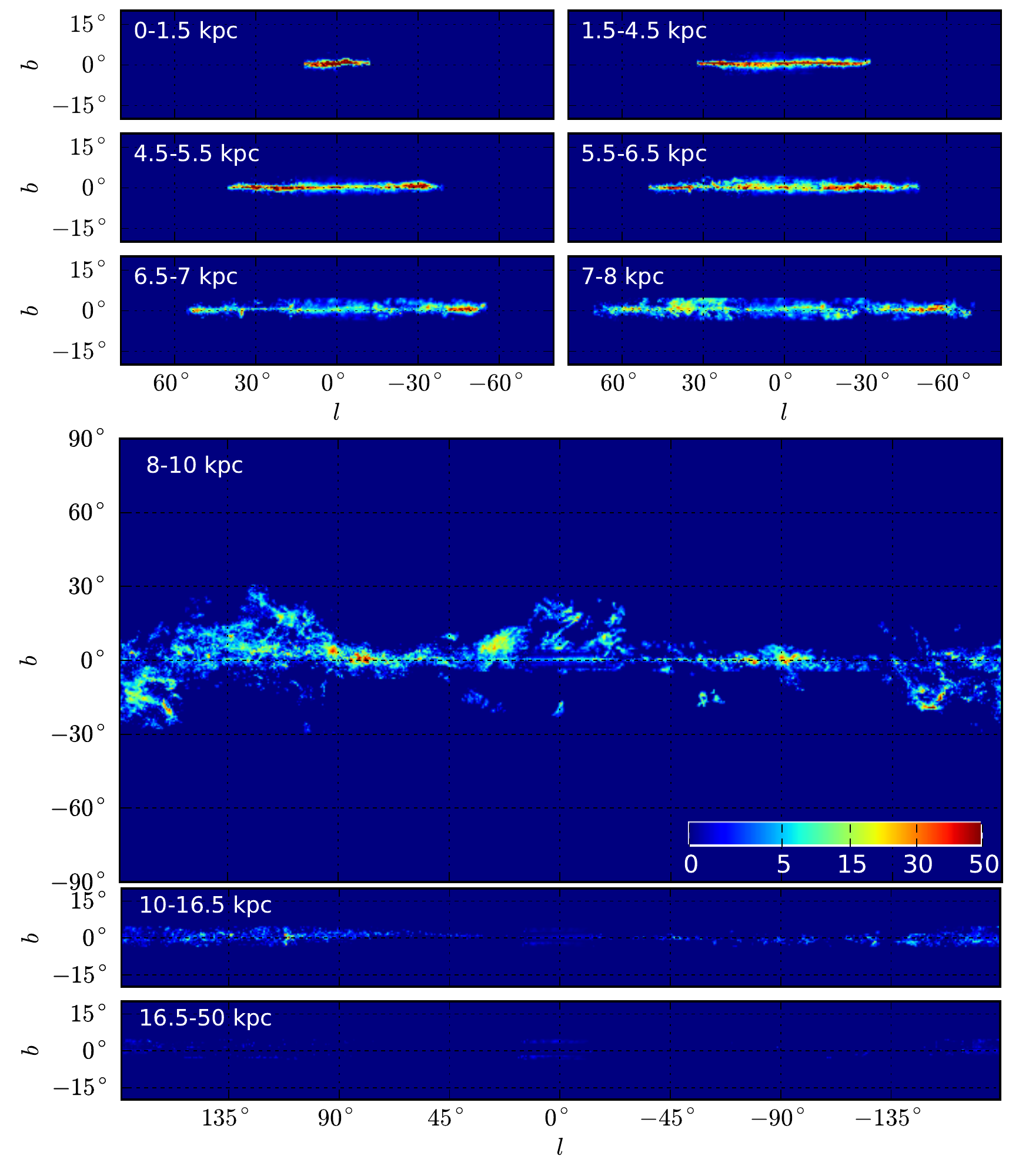}
%  \caption{$W$(CO)}
%%%%  \label{co_rings}
\caption{Galactocentric annuli of $N_{\text{H}\textsc{~i}}$ in 10$^{20}$~cm$^{-2}$ (left) and $W$(CO) in K~km~s$^{-1}$ (right), displayed in Galactic plate carr\'ee projection with bin size of 0\fdg125 $\times$ 0\fdg125.  
The square root color scaling saturates at 100$\times$10$^{20}$~cm$^{-2}$ for \nhi and at 50~K~km~s$^{-1}$ for $W$(CO). The Galactocentric boundaries for each annulus are written in each panel.}
  \label{hi_rings}
\end{figure}

Because our model assumes a uniform CR density in each template, it is crucial to partition the Galaxy into Galactocentric annuli to account for radial variations in the CR density. The radial velocities measured from the Doppler shifts of the 21-cm line radiation can provide kinematic Galactocentric distances of \hi clouds. We assume that the gas moves in circular orbits around the Galactic Center (GC) and use the rotation curve given by \cite{Clemens:1985p914} with a Galactocentric distance and speed of the Sun of 8.5~kpc and 220 km~s$^{-1}$ respectively. Measurement of the Galactic rotation curve from 21-cm and 2.6-mm surveys was a well understood procedure by the 1980s, for the inner Galaxy primarily involving the measurement of the terminal velocity as a function of longitude. The Milky Way is a barred spiral, and a rotation curve does not strictly apply within Galactocentric radius range of the bar.  However, the relatively coarse binning in radius that our model fitting requires mitigates the effect of noncircular motions in a barred potential. Regions located within 10$\degr$ of the GC and anti-center have poor kinematic distance resolution. We have linearly interpolated the column densities in Galactic longitude from the column density integrated within 5$\degr$ of longitude from their boundaries and scaled each line of sight to the total \nhi. The innermost \hi annulus, where this interpolation is not possible, is assumed to contain 60\% more gas than its neighboring annulus. We have excised the nearby galaxies LMC, SMC, M33 and M31, which were within the velocity range of the LAB survey. The %Galactocentric derivation and
partitioning of atomic hydrogen and CO into annuli is detailed in Appendix B of \cite{Ackermann:2012p2978}. We have generated 9 annuli (see Figure \ref{hi_rings}) with limits given in Table \ref{rings}.  Annulus number 7,  called ``local'' spans the solar circle. For each annulus, since the scale height of the CR distribution is several times greater than that of the gas, we assumed a uniform density in the axes perpendicular to the Galaxy.

\begin{deluxetable}{rrr}
\tabletypesize{\scriptsize}
\tablecaption{Limits of the $N_{\text{H}\textsc{~i}}$ Galactocentric annuli in kpc.\label{rings}}
\tablewidth{0pt}
\tablehead{\colhead{Annulus}&\colhead{$r_{min}$ (kpc)}&\colhead{$r_{max}$ (kpc)}}
\startdata
        1 & 0.0 & 1.5 \\
        2 & 1.5 & 4.5 \\
        3 & 4.5 & 5.5 \\
        4 & 5.5 & 6.5 \\
        5 & 6.5 & 7.0 \\
        6 & 7.0 & 8.0 \\
        7 & 8.0 & 10.0 \\
        8 & 10.0 & 16.5 \\
        9 & 16.5 & 50.0 \\
\enddata
%\tablecomments{Differential emissivity}
\end{deluxetable}

%\begin{figure}
%    \subfloat[$N_{\text{H}\textsc{~i}}$ \label{hi_rings}]{%
%      \includegraphics[width=0.48\textwidth]{hi_rings}
%    }
%    \hfill
%    \subfloat[$W$(CO) \label{co_rings}]{%
%      \includegraphics[width=0.48\textwidth]{co_rings}
%    }
%	\caption{Galactocentric annuli of $N_{\text{H}\textsc{~i}}$ in 10$^{20}$~cm$^{-2}$ and $W$(CO) in K~km~s$^{-1}$ displayed in Galactic plate carr\'ee projection with bin size of 0\fdg125 $\times$ 0\fdg125.  
%For all the panels we applied a square root scaling, the maximum of the scale corresponding to 100$\times$10$^{20}$~cm$^{-2}$ for $N_{\text{H}\textsc{~i}}$ and to 50~K~km~s$^{-1}$ for $W$(CO). The Galactocentric boundaries for each annulus are written in each panel.}
%    \label{fig:rings}
%  \end{figure}

%\begin{subfigure}{.5\textwidth}
%  \centering
%  \includegraphics[width=8.4cm]{hi_rings}
%  \caption{$N_{\text{H}\textsc{~i}}$}
%  \label{hi_rings}
%\end{subfigure}%
%\begin{subfigure}{.5\textwidth}
%  \centering
%  \includegraphics[width=8.4cm]{co_rings}
%  \caption{$W$(CO)}
%  \label{co_rings}
%\end{subfigure}
%\caption{Galactocentric annuli of $N_{\text{H}\textsc{~i}}$ in 10$^{20}$~cm$^{-2}$ and $W$(CO) in K~km~s$^{-1}$ displayed in Galactic plate carr\'ee projection with bin size of 0\fdg125 $\times$ 0\fdg125.  
%For all the panels we applied a square root scaling, the maximum of the scale corresponding to 100$\times$10$^{20}$~cm$^{-2}$ for $N_{\text{H}\textsc{~i}}$ and to 50~K~km~s$^{-1}$ for $W$(CO). The Galactocentric boundaries for each annulus are written in each panel.}
%\end{figure}

\subsubsection{Molecular  hydrogen}
The molecule H$_2$, which does not have a permanent dipole moment, cannot be observed in direct emission in its dominantly cold phase. The observation of molecular gas relies on other molecules and especially on the 2.6-mm J=1$\rightarrow$0 line of $^{12}$C monoxide (CO), the second most abundant molecule in the ISM. The millimeter-wave emission of CO can trace H$_2$ because the molecular hydrogen is its main collisional partner, and collisions excite %and de-excite %
its rotational transitions \citep{Wilson:1970p4389,Yang:2010p4065}. Despite the large optical thickness of the low-level rotational lines of CO, numerous studies suggest the H$_2$ column density to be proportional to the velocity-integrated CO brightness temperature $W$(CO). This relation was experimentally observed by comparing the virial masses of various molecular clouds to their CO luminosities, and interpreted by \cite{Solomon:1987p4110} who inferred that molecular clouds have a rich substructure of small optically thick regions of distinct velocity, conceptually analogous to a mist made of discrete droplets. This molecular `mist' is optically thin at each velocity and the CO line intensity is proportional to the total number of molecular `droplets' along the line of sight. The molecular-hydrogen-to-CO conversion factor (an assumed proportionality between the integrated column density in the CO line and the column density of H$_2$) is expressed as $X_{\text{CO}}=N_{\text{H}_{2}}$/$W$(CO). We have obtained the $W$(CO) spatial distribution from the Center for Astrophysics composite survey \citep{Dame:2001p1849}. It is composed of a Galactic plane survey with a sampling interval of 0\fdg125 and surveys covering all large local clouds at higher latitudes with a sampling interval of $0\fdg25$. We have also used dedicated observations lying outside the sampling boundary of the composite CO survey at northern declinations (T. Dame 2011, private communication). This CO survey covers the great majority of the Galactic CO emission \citep{Ade:2014p4070}. We have derived the integrated intensities using a ``moment mask'' filtering to enhance the signal-to-noise ratio \citep{Dame:2011p4025}. We have derived Galactocentric annuli from radial CO velocities in the same way as for \hi (Figure \ref{hi_rings}) in order to allow for variations in CR density, but also in \xco. The innermost CO annulus contains all high-velocity CO emission. 
%To lower the number of templates in the foreground of
%improve the spatial contrast between $N_{\text{H}\textsc{~i}}$ and $W$(CO) in 
%the inner Galaxy we combined annuli 6 and 7 (Table \ref{rings}) into a wider ring. 
We have combined the outer annulus 8 with the annulus 9, which lacks measurable CO emission to be reliably fitted. This is equivalent to assuming that the CO is immersed in a uniform CR flux, with a constant \xco, between 10 and 50~kpc. Parts of the Aquila Rift molecular cloud were incorrectly attributed to the CO annulus 6 by this procedure; we have also merged this annulus with CO annulus 7.  

The central molecular zone (CMZ) is a massive complex of giant molecular clouds located in the central region of the Galaxy \citep{Serabyn:1996p4400,Ferriere:2007p1322}. It appears to be pervaded by intense magnetic fields \citep{Morris:2014p4398} and recently hosted a burst of star formation \citep{YusefZadeh:2009p4404}. Because of these unique conditions, we have cut out the CMZ from our innermost ring and created a dedicated CMZ column-density map in \hi and CO. For that we have selected a contour corresponding to 20~K~km~s$^{-1}$ in W(CO) for longitude $-1\fdg5<l<4\fdg5$ and we have assigned all the molecular gas inside the contour to the CMZ. We have used the same contour to extract from the LAB survey an \nhi map for the CMZ.

\subsubsection{Ionized hydrogen}
There is no direct observational information on the %accurate representation of 
spatial distribution of the warm ionized medium. \cite{He:2013p4087} studied 68 radio pulsars detected at X-ray energies and compared the free electron column density given by dispersion measures to \nhi along the line of sight as traced by X-ray extinction; they obtained a ratio of H$^{+}$ to \hi column density in the range of 0.07 to 0.14. The H-$\alpha$ emission, a two-particle process proportional to the integral of the square of the electron density along the line of sight, suffers from dust absorption at low latitude \citep{Dickinson:2003p4084} and from scattering by interstellar dust \citep{Witt:2010p4082, Seon:2012p4083}. The free-free emission, also proportional to the square of the electron density, is not absorbed in the radio, but difficult to separate from the synchrotron emission and has contributions from numerous H~{\sc ii} regions requiring careful temperature correction. \cite{Cordes:2002p3302} developed a model (NE2001) of the density distribution of Galactic free electrons based on 1143 dispersion measures of pulsars with known distances. We have built column-density annuli maps for the warm ionized medium based on NE2001 predictions, but adding this component to the \g-ray model did not improve the fit to the LAT data. We performed several tests like excluding the Galactic plane from the fit or removing individual NE2001 clumps of ionized gas that seemed over predicted but we were not able to improve the fit likelihood. \cite{Paladini:2007p813} also found that the fit of infrared dust emission worsened with an H$^{+}$ column-density map ($N_{\text{H}^{+}}$) extracted from NE2001, probably because of its simplified spatial distribution. Figure 7 of \cite{Sun:2008p3774} shows that NE2001 does not reproduce the structures of the free-free emission from the Wilkinson Microwave Anisotropy Probe (WMAP) observations. Based on \cite{Gaensler:2008p3776}, we have used an exponential scale-height of 1~kpc to build a simple $N_{\text{H}^{+}}$ map, which also failed to improve the fit to the LAT data. 
The H$^{+}$ template normalization was set to zero by the fit and no H$^+$ related structure appeared in the final residual \g-ray map. With the present sensitivity of the LAT survey we did not detect \g rays specifically originating from the small mass in the diffuse H$^{+}$ layer and we have dropped the $N_{\text{H}^{+}}$ map from the models.
%are probably accounted for by other templates. The final model accounts for the $\gamma$ rays originating from H$^{+}$ but does not contain any specific $N_{\text{H}^{+}}$ template. 

\subsubsection{Dark neutral medium}
\label{sec:Dark_neutral_medium}
The $^{12}$CO J($1\rightarrow 0$) line emission is not a perfect tracer of cold H$_2$. In addition to metallicity variations, the CO molecule is strongly affected by UV photodissociation in the outer regions of molecular clouds where H$_{2}$ can exist without CO or where CO is only weakly excited \citep{Wolfire:2010p3814, Glover:2011p3778}. \xco also depends on the dynamical characteristics of the molecular cloud. \cite{Shetty:2011p4064} went beyond the simple ``mist'' model and calculated the radiative transfer of the CO line in turbulent clouds in order to investigate the dependence of \xco on the physical properties of gases. They predicted that the typical ranges of mean column density, temperature, and velocity dispersion found in molecular clouds lead to a variation in \xco by about a factor of 2. Moreover, as we already mentioned, the column density of \hi derived from the 21-cm brightness temperature under the hypothesis of a uniform $T_{S}=140$~K is likely to be biased in lines of sight for which $T_{S}$ is not uniformly 140~K. 
Those limitations lead to large underestimates of the quantities of gas at the \hi--H$_2$ interface in our Galaxy \citep{Grenier:2005p836, Ade:2011p3178, Paradis:2012p4067} as well as in the Magellanic Clouds \citep{Bernard:2008p4069, Dobashi:2008p1120}. 
This transitional region is referred to as the dark neutral medium (DNM) and for any particular region it comprises unknown fractions of cold dense \hi and CO-free or CO-quiet H$_2$. 

Neutral gas and large dust grains coexist; observations have shown that the gas-to-dust ratio leads to a mass ratio $M_{gas}/M_{dust}\sim 100$. Dust column density can therefore provide an alternative template for \hi and H$_2$ in the Milky Way. We used the dust reddening map of \cite{Schlegel:1998p290}, which is based on the 6$\arcmin$ resolution IRAS/ISSA emission map at 100~$\micron$ 
%that depends on dust column density and also on the dust temperature. This temperature varies with the optical properties and size of the dust, and with the interstellar radiation field intensity. \cite{Schlegel:1998p290} corrected the infrared (IR) emission map with the dust temperature measured from the ratio of the accurately calibrated COBE/DIRBE at 100 and 240 $\micron$. The resulting map, once calibrated to elliptical galaxies, 
and on a correction for the dust temperature inferred from the emission ratio measured between 100 $\micron$ and 240 $\micron$ at 0\fdg7 resolution with COBE and DIRBE. The resulting map, once scaled to reddening to match the data from elliptical galaxies, has often been used
as an estimator of Galactic extinction. It has been recently superseded by dust optical depth derivations at higher angular resolution and with a broader spectral coverage thanks to the Planck data \citep{Abergel:2013p4126}, but those were not available for the development of the present GIEM for \textit{Fermi}-LAT. 

\begin{figure}
\begin{center}
\includegraphics[width=12cm]{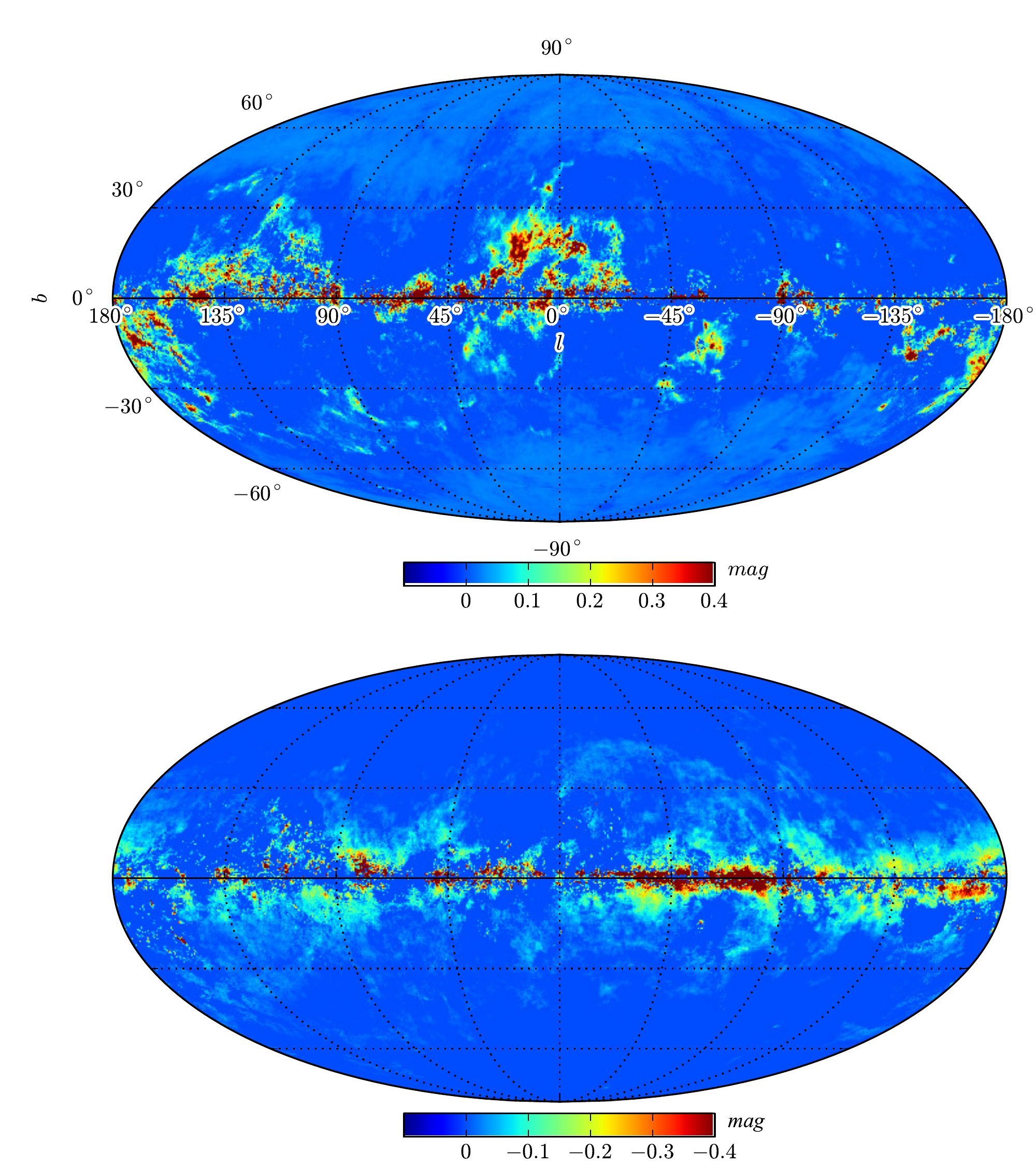}
\caption{Mollweide projection in Galactic coordinates of the excess (top) and deficit (bottom) of the dust reddening E(B-V) over the best linear combination of \hi column-density and W(CO) intensity maps. Assuming a uniform dust-to-gas ratio and a uniform dust grain radiative emissivity, the dust reddening map can be used as template for the total gas column density. The excess map predominantly traces the DNM column density of dense \hi and diffuse H$_2$ gas. The strong deficits ($< -0.2$ mag) along the Galactic plane are interpreted as $N_{\text{H}\textsc{~i}}$ corrections in regions where $T_{S}>140$~K. The smaller deficits that extend to 10$\degr$ or 20$\degr$ in latitude are likely due to grain emissivity variations in the warm dust of diffuse cloud envelopes. } 
%It may also incorporate H$^{+}$ and a correction over the hypothesis of a uniform $T_{S}=140$~K. 
%To trace the dust thermal emission we used the dust optical depth map of \cite{Schlegel:1998p290} normalized as E(B--V) reddening expressed in mag.}
\label{EBV}
\end{center}
\end{figure}

Away from photo-dissociation and hot star-forming regions where dust properties can vary dramatically,
%cold dust is assumed to be well mixed with the gas and thus provides a good template for the total gas, including the DNM not traced by \hi and CO  emission. 
most of the dust column density is well correlated with $N_{\text{H}\textsc{~i}}$ and $W$(CO). Regions of dust not correleted with $N_{\text{H}\textsc{~i}}$ and $W$(CO), that spatially correlate with diffuse $\gamma$-ray excesses over \hi and CO expectations, correspond to DNM-rich clouds \citep{Grenier:2005p836}. 
We have derived residual maps by subtracting from the dust reddening map the parts linearly correlated with the $N_{\text{H}\textsc{~i}}$ and $W$(CO) annuli. We have first filtered out IR point sources present in the dust map and applied inpainting methods based on wavelet decomposition \citep{Elad:2005p4149} to reconstruct the signal at the point-source locations. We have then proceeded in steps in order to minimize the impact of the model uncertainties in the inner Galaxy onto the closer annuli: we first fitted the reddening map to the local gas annuli for $|b| > 10\degr$; then we fixed them and fitted the outer annuli 8 and 9 in the anticenter region ($90\degr<l<270\degr$); then we fixed the previous annuli and fitted the inner Galaxy. Subtracting the correlated parts from the total dust has revealed coherent structures across the sky in both the positive and negative residuals 
%($-$0.2 to 0 magnitude, typically) and other clouds with more pronounced positive residuals (0 to 0.6 mag). 
(see Figure \ref{EBV}). We note that when we fitted the \hi maps to derive the $T_{S}$ that gives the best fit to the {\it Fermi} data (Figure \ref{TS}), we have used corresponding $T_{S}$ values to extract the dust residual maps.  %We accounted for this gas by including in our model a template for total dust column density.

The residuals shown in Figure \ref{EBV} are statistically significant above 0.04~mag. The positive residuals reveal gas in addition to that traced by $N_{\text{H}\textsc{~i}}$ and $W$(CO). We associate this excess map to the DNM distribution. For a standard $N_{{\rm H}}/E(B-V)$ ratio \citep{Bohlin:1978p2105,Casandjian:2015p4380}, they translate to column-densities in excess of $10^{21}$ cm$^{-2}$ in nearby clouds and in more distant regions of the Galactic disc. There is no velocity information associated with the detection of IR dust emission, so we could not partition the DNM into annuli. This means that all DNM clouds effectively have the same CR flux in the $\gamma$-ray model fitting. Dense \hi and diffuse H$_2$ dominate the DNM column-densities, but the residual map also potentially incorporates ionized hydrogen mixed with dust. The ionized mass is, however, small compared to the DNM one, which is known in nearby clouds to compare with or exceed the mass locked in the CO-bright H$_2$ cores \citep{Grenier:2005p836, Ade:2011p3178, Ade:2014p4235}.
%even if it also includes regions in which an average $T_{S}$ of 140~K is too high and potentially incorporates also ionized hydrogen that might also be mixed with dust. 
% that resides predominantly at the interface between the \hi and CO structures \citep{Grenier:2005p836}. 
From the DNM map, we have excised the Magellanic Clouds, M31, as well as regions with a high density of IRAS sources including the inner Galaxy for absolute longitudes less than 30$\degr$ and latitudes less than 2$\degr$, and part of the Cygnus region. We have also inpainted the excised regions using the same method mentioned above. 
%This gas can be cold dense \hi not accounted for in the N(H~I) maps in regions where the spin temperature is closer to 30 or 40~K \citep{Heiles:2003p1829}. It is also likely to include moderately dense H$_2$ that is known to exist in abundance at the interface between the atomic and molecular phases, but that is not dense enough ($< 10^3$ cm$^{-3}$) to efficiently excite CO lines, typically for E(B-V) $<$ 0.6 mag \citep{Lombardi:2006p2001,Leroy:2009p2995}. 
%dust component represents about 1\% of the total ISM mass dust component 

The negative residual map in Figure \ref{EBV} exhibits strong deficits close to the Galactic plane where the \hi and CO expectations exceed the data. They  
are likely related to regions in which an average $T_{S}$ of 140~K is too low, and thus $N_{\text{H}\textsc{~i}}$ is overestimated. The inclusion of these potential corrections to \nhi in the $\gamma$-ray model does improve the fits to the interstellar $\gamma$-ray emission. In this paper we call this map the ``$N_{\text{H}\textsc{~i}}$ correction map''. We also observe strong negative residuals in regions close to bright OB associations where the dust temperature corrections are too coarse because of the low spatial resolution of DIRBE. Such residuals in Orion have led to 
the detection of spurious $\gamma$-ray point sources in the 2FGL source catalog (flagged as 'c' sources). We have zeroed those residuals in the Orion molecular cloud.
We also observe small, diffuse residuals that extend to about 20$\degr$ in latitude. They correspond to regions of warm dust, with temperatures of 18--19~K and soft emission spectra ($\beta < 1.5$) in the recent spectral studies joining millimeter and IR data from Planck and IRAS \citep{Abergel:2013p4126}. They often surround regions of strong PAH emission or free-free emission, thereby suggesting dust exposed to a different interstellar radiation field than in the \hi and CO clouds and for which the temperature corrections applied by \cite{Schlegel:1998p290} were not precise enough. These residuals are thus more likely to reflect emissivity changes of the dust grains rather than corrections to the total gas column densities. 
%We did not attempt to remove them because they would translate to small decreases in column densities 
%%($< 10^{21}$ cm$^{-2}$) 
%compared to the values present in the DNM or in the \hi and CO annuli of Figures \ref{hi_rings}. 
%%We applied a flux density of 300~Jy in the Iras map at 100~$\mu$

%The positive residuals associated with the DNM, $T_{S}$ correction and potentially the ionized hydrogen are shown in the upper part of the Figure \ref{EBV} in units of magnitude. In the lower part of the same figure we show the $N_{\text{H}\textsc{~i}}$ correction map. 
%We multiplied the maps by an arbitrary gas-to-dust ratio to obtain gas maps in column density. 

In the local ISM, \cite{Grenier:2005p836} have shown that the use of \hi, CO, and a dust residual map improves the fits to the interstellar $\gamma$-ray emission 
%There is in principle no benefit from using HI, CO, and a dust residual map 
compared to using the sole dust map for the total gas.
While the hydrogen is likely to have the same $\gamma$-ray emissivity per atom in the atomic, DNM, and molecular components of the ISM because of the good penetration of CRs at the GeV-TeV energies relevant for the LAT \citep{Skilling:1976p4058}, dust opacity (optical depth per gas nucleon) is known to increase from the diffuse \hi  to dense \hi and to H$_2$ gas \citep{Stepnik:2003p4118,Abergel:2013p4126,Abergel:2014p4401,Ade:2014p4235}. Restricting the use of dust to tracing the DNM alleviates the impact of dust opacity gradients because the DNM spans a moderate range of gas volume densities and is rather diffuse in space. Using a unique dust column-density map at high latitude instead of \hi, CO, and dust-inferred DNM maps would also lead to discontinuity in the model close to the Galactic plane where \hi and CO annuli must be used to follow CR variations with distance from the GC. 

%Additionally this extra step results in a DNM map that has spatial structures of smaller angular scales than the more diffuse dust optical depth map. Those small-scale structures improve the wavelet decomposition for the inpainting of excised regions. \isa{???} 

%Note that we assumed here an uniform $T_{S}$ and corrected this approximation by fitting both the positive and negative residuals of the dark gas map. The $T_{S}$ of 140~K is therefore the temperature that best fit the outer Galaxy under the assumption that the dust column density of \cite{Schlegel:1998p290} is accurate and that the dust is mixed with all the phases of \hi.

\subsection{Galactic Inverse Compton radiation}
\label{sec:Galactic_Inverse_Compton_radiation}
%The calculation takes into account propagation and all energy losses and gains, with the IC energy losses for the cosmic-ray electrons and positrons (including Klein-Nishina effects) calculated using the full spatial and energy distribution of the ISRF.
While the different gas column-density maps offer spatial templates for $\gamma$-ray photons originating mainly from $\pi^{0}$-decay and bremsstrahlung emission, there is no direct observational template for the IC emission. Instead it must be calculated. We have used the prediction from the GALPROP code with GALDEF identification $^SY^Z6^R30^T150^C2$. This model features a radial distribution of CR sources proportional to the distribution of pulsars in the Galaxy given by \cite{Yusifov:2004p3648}, a Galactic halo size and radial boundary equal to 6~kpc and 30~kpc, respectively, and a representative diffusive reacceleration model described in \cite{Ackermann:2012p2978}. In that work, the GALPROP code was run to obtain models for the primary and secondary CR electron and positron intensities and spectra throughout the Galaxy. GALPROP then folds the distributions with the ISRF \citep{Porter:2008p3784} to obtain the IC emissivity, which was integrated along the line-of-sight for each direction and energy range to obtain IC intensity sky maps. We use $I_{IC_{p}}$ to denote the intensity of the predicted IC emission.

\subsection{\loopI and \fb}
\label{sec:Large_scale_structures}

\begin{figure}
\begin{center}
\includegraphics[width=16cm]{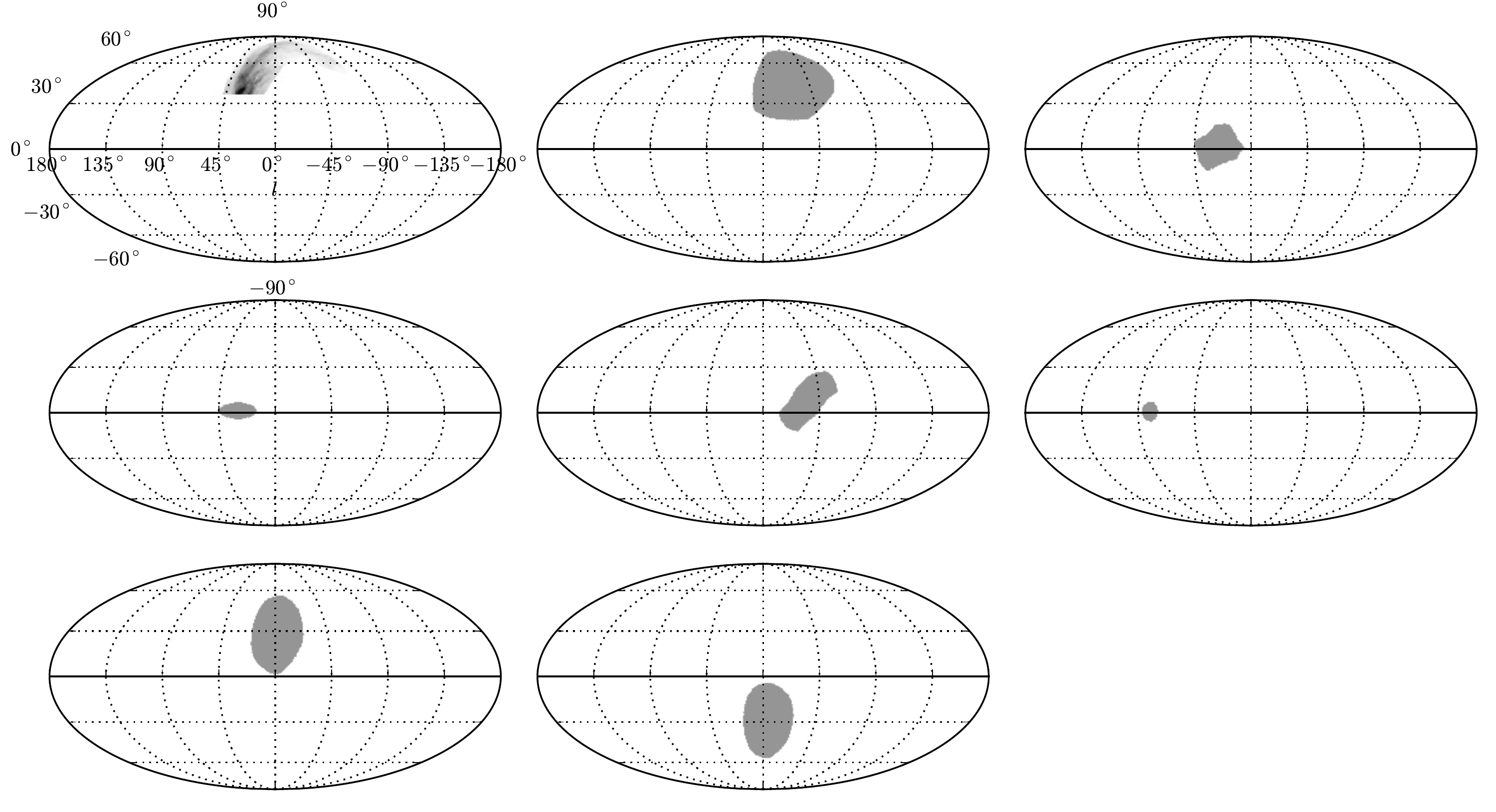}
\caption{Intensity distributions, in Galactic coordinates and arbitrary units, used for the North Polar Spur (from the 408~MHz map) and for uniform patches added to account for regions of Extended Excess Emissions while deriving the gas emissivities in the nearby and outer annuli (see sections \ref{sec:fitting_procedure_section} and \ref{sec:local_annuli}). The last two patches correspond to the \fb and the previous one to the region around the cocoon of hard spectrum cosmic rays in Cygnus X. Those patches were not used in the final interstellar model.} 
%without bias from unaccounted excesses in the fit to {\it Fermi}-LAT counts map we modeled $\gamma$-ray excesses associated with the North Polar Spur with a template derived from radio synchrotron at 408~MHz. The other excesses were accounted for by patches with shapes of uniform intensity. None of these templates {\bf was used to derive the inner or outer Galaxy annuli emissivities nor } was incorporated into the final interstellar emission model.}
\label{fig_patches}
\end{center}
\end{figure}

%The template method assumes that the CR flux is uniform within each template. Since the CR density varies with Galactocentric distance we partitioned the Galactic disk into annuli, but given that the gas distribution is not uniform within a Galactocentric annulus the CR flux is probably not uniform either. We used the same approximation for the IC$_{p}$ intensity calculated by GALPROP, for which the calculated distribution of CR electrons is cylindrically symmetric. This approximation is the most likely reason for excesses observed when we compared a preliminary template model derived only from the templates mentioned above to the LAT observations.
%There is for example a large excess of $\gamma$ rays probably coming from a population of CR electrons trapped in the \loopI giant radio continuum loop
The sky at GeV energies also features large angular scale structures such as \loopI and the \fb. In the radio loops, the $\gamma$ rays are likely produced by a population of CR electrons trapped in the old supernova remnants \citep{Grenier:2005p4078, Casandjian:2009p3311, Ackermann:2012p2978}. The \fb correspond to two lobes of hard emission, extending North and South from the direction of the GC \citep{Su:2010p2675, Ackermann:2014p4189}. There is no accurate independent template for the $\gamma$-ray emission of those large structures at other wavelengths since the synchrotron maps available in the radio fold the CR electron distribution with the complex structure of the magnetic field, and X-ray maps trace the hot gas, but not the high-energy particle content. 
In the first iteration of our fitting we assumed a proportionality between the $\gamma$-ray intensity and the bright radio continuum emission of the North Polar Spur (NPS), which dominates 
%Not including them in Equation \ref{eqRing}, as well as other structures that we observed in the residuals at lower latitudes, will strongly bias the fit. To reduce this bias we roughly modeled the North Polar Spur, the bright \loopI radio filament observed at 30$\degr$ in Galactic longitude, 
the 408~MHz radio continuum intensity survey of \cite{Haslam:1982p4092} at large positive latitudes. To do so, we have subtracted the point sources from the radio map and selected a region around \loopI (first panel of Figure \ref{fig_patches}). In the absence of external maps of the \fb, we have assumed a uniform intensity template in the first iteration of our fitting (see Section \ref{sec:fitting_procedure_section} and Figure \ref{fig_patches}) and for the final model we have extracted intensity maps from the LAT data (see Section \ref{extra_emission}).

\begin{comment}  % !!!!!!!!!! commented

%% Table: Patches
\begin{deluxetable}{lccccc}
\tabletypesize{\scriptsize}
\tablecaption{Additional Components in the Diffuse Emission Model
\label{tbl:patches}}
\tablewidth{0pt}
\tablehead{\colhead{Designation}&\colhead{Center}&\colhead{Approx. dim.}&\colhead{ $\Omega/ 4\pi$}&\colhead{Fraction of total}&
\colhead{Fraction of intensity} \\\colhead{}&\colhead{$(l, b)$}&\colhead{$(l \times b)$}&\colhead{}&\colhead{intensity}&\colhead{within patch}}
\startdata
First quadrant and inner   &  $25\degr, 0\degr$  &   $40\degr \times 30\degr$   & 1.9\% & 1.0\% & 13.4\% \\                                                          
Fourth quadrant                 &   $-$35, 9                  &    $40\times30$                      &   2.4 &; 0.3 & 3.8 \\
Lobe North                          &     0, 25                      &    $50\times40$                      &   3.9 & 0.4 & 6.9 \\
Lobe South                         &      0,$-$30                &      $50\times40$                     &   3.7 &; 0.4 & 14.1 \\
\enddata
\tablecomments{Description of the additional components added in the Galactic diffuse model. The centers and extents are in Galactic coordinates.  The extents are approximate because the shapes are irregular.  $\Omega$ is the solid angle.  To evaluate the fractional intensities, we integrated the intensity above 130~MeV for the First and Fourth quadrant patches and above 1.6~GeV for the Lobes patches.}
\end{deluxetable}

\end{comment}   % !!!!!!!!!! commented

\subsection{Point sources}
\label{section_pt_sources}
We have modeled each of the 2179 point sources derived with a preliminary iteration of the GIEM for the 3FGL catalog \citep{Ballet:2013p4107,Acero:2015p4385}. We have modeled each source with the Science Tool {\it gtsrcmaps} \footnote{\label{fssc}The {\it Fermi} Science Tools analysis package, the LAT $\gamma$-ray data, and the Instrument Response Functions (IRFs) are available from the {\it Fermi} Science Support Center, \url{http://fermi.gsfc.nasa.gov/fssc}} that takes into account the exposure, the angular resolution, and the source spectrum at each source position and in each energy bin. 

%In order to check the fit procedure we verified that the spectral energy distribution (SED) obtained by the fit for each source follows typical spectral forms expected for $\gamma$-ray sources. We fitted each source SED between 50 MeV and 50 GeV with three spectral identical to the ones used in \cite{Nolan:2012p3051}: a simple power-law, an exponentially cutoff power-law, and a log-normal. Figure \ref{fig_chi2_src} presents a histogram of the reduced $\chi^2$ resulting from this fit, using for each of the 2179 sources the lowest $\chi^2$ among the three cases. We observe that 68\% of the sources have a reduced reduced $\chi^2$ of less than 4 which is reasonable given our large energy range and the limited number of spectral forms tested for the point-source SEDs. % with simple spectral forms. 
%We did not use these fits or spectral representations to derive any component of the interstellar model.
%In Equation \ref{eqRing} we did not explicitly include a term corresponding to the contribution of unresolved (undetected) Galactic sources.

\cite{Ackermann:2013p4156} analyzed the source population to estimate the contribution of unresolved sources to the diffuse Galactic emission above 10~GeV, and found the contribution to be about 2.5\% of the interstellar emission for a reference model with a local source density of 3~kpc${^{-3}}$. For a tenfold increase in the local source density in a ``maximum density'' model, their contribution rises to ~8\%. In \cite{Acero:2015p4385}, it was estimated that their contribution at 1~GeV amounts to 3\% in the inner Galaxy. In the present work, the $\gamma$ rays produced by unresolved sources are likely accounted for by the other templates, in particular the IC and inner \hi templates.

%\begin{figure}
%\begin{center}
%\includegraphics[width=12cm]{chi2_src}
%\caption{Histogram of the reduced $\chi^2$ of the SED fits for 2179 point sources, based on their best-fitting spectral forms, either a power-law, an exponentially cutoff power-law, and a log-normal.}
%\label{fig_chi2_src}
%\end{center}
%\end{figure}

\subsection{Extended and moving sources}
We have built intensity maps for the following 21 extended sources:  Centaurus A, Cygnus Loop, Gamma Cygni, HESS J1614$-$518, HESS J1616$-$508, HESS J1632$-$478, HESS J1825$-$137, HESS J1837$-$069, IC 443, LMC, MSH 15-52, Puppis A, RX J1713.7$-$3946, S147, SMC, Vela Junior, Vela X, W28, W30, W44, and W51C. These small extended sources are associated with specific supernova remnants, pulsar wind nebulae and spatially-resolved galaxies. We have modeled each source with a simple disk shape, a ring, a 2D Gaussian function, or a map derived from other wavelengths, as described in \cite{Nolan:2012p3051} and \cite{Ballet:2013p4107}. 

The time-integrated $\gamma$-ray emissions from the Sun and the Moon effectively add a diffuse glow across the sky, at low ecliptic latitudes. Their intensities and spatial distributions have been calculated by \cite{Abdo:2011p4076, Abdo:2012p4077} and used here.

\subsection{Isotropic intensity}
\label{sec:isotropic_section}
The isotropic emission encompasses the isotropic diffuse $\gamma$-ray background \citep{Ackermann:2015p4239, Abdo:2010p4079} %essentially 
originating from unresolved sources like blazars, star-forming and radio galaxies as well as contamination from the very small fraction of CRs interacting in the LAT that are misclassified as $\gamma$ rays and from Earth limb photons that enter the LAT from the back but are reconstructed to have come from within the field of view. Because of this contamination, the isotropic emission depends on the event class and the conversion type in the LAT \citep{Ackermann:2012p4094}. We have modeled this emission with an isotropic intensity template, with an intensity spectrum to be obtained from the fit to the $\gamma$-ray in each energy band.

\subsection{Residual Earth limb emission}
\label{sec:limb_section}
\begin{figure}
\begin{center}
\includegraphics[width=12cm]{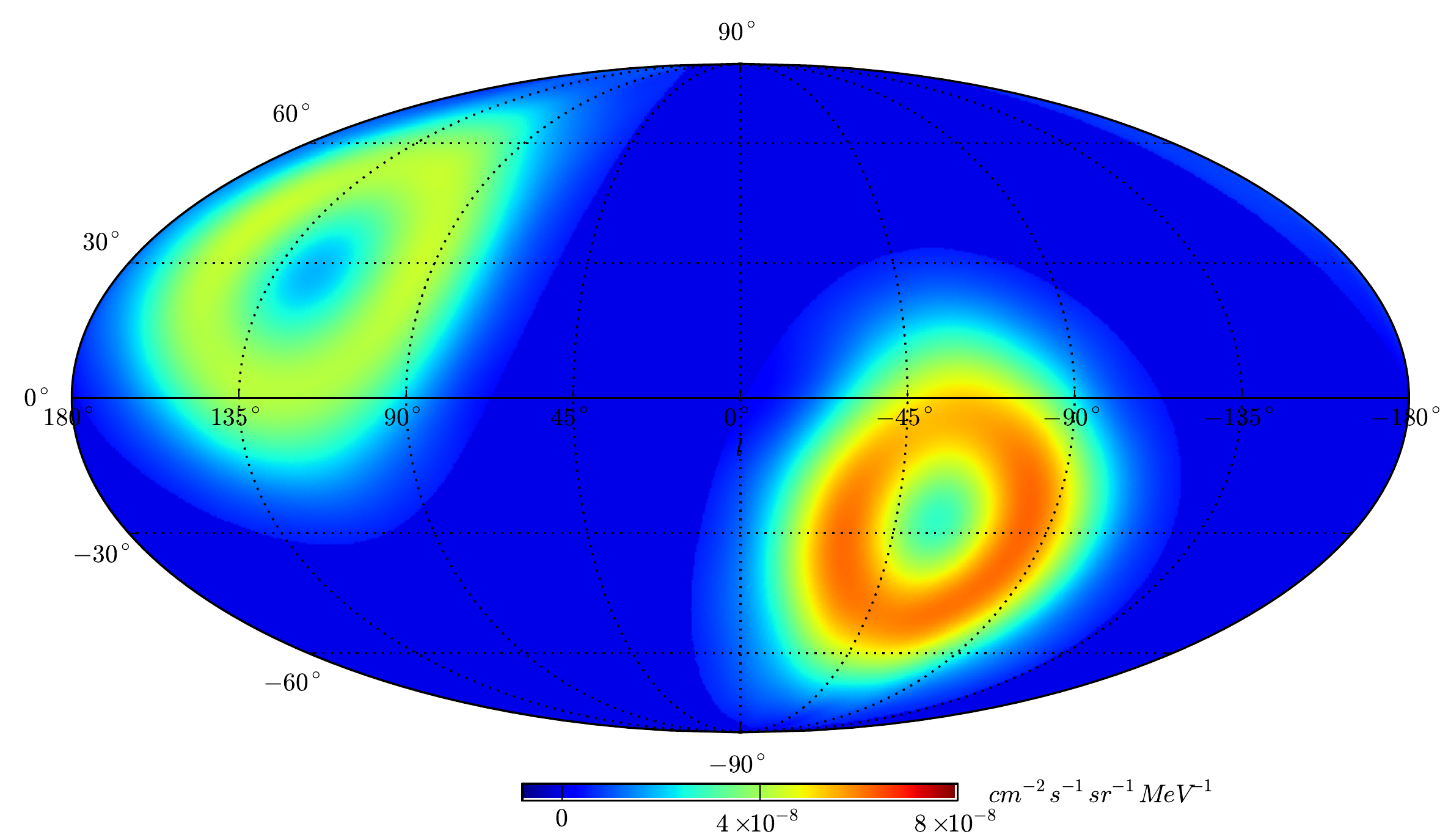}
\caption{All-sky distribution, in Galactic coordinates and photon intensity, of the residual emission originating from the Earth limb at 100 MeV.}
%Shape and intensity of the Earth limb template at 100~MeV.}
\label{limb}
\end{center}
\end{figure}

Due to their proximity, CR protons and electrons interacting with the Earth's atmosphere make it by far the brightest $\gamma$-ray source in the sky, with intensity $\sim$1000 times larger than that of the Galactic plane \citep{Abdo:2009p4150}.  The {\it Fermi} standard observational strategy is such that the Earth is not directly in the field of view of the LAT.  However, a large number of limb photons entering the LAT from the side are still detected. We have removed the great majority of those photons with a simple cut in the zenith angle at 100$\degr$, but a residual contamination coming from the tails of the PSF is still observed at energies below about 200~MeV. Accurately simulating this component of the Earth limb photons in the far tails of the PSF is challenging, so we have chosen instead to construct a simple template based on the subtraction of the map derived with a zenith angle cut at 80$\degr$ from one restricted to angles above 100$\degr$ for energies between 40~MeV and 80~MeV. We have deconvolved the resulting map to account for the PSF broadening. We used the assumption that the spatial distribution of residual limb photons, which over the long time interval analyzed here is largely determined by the inclination and altitude of the orbit, is independent of energy. The all-sky distribution is displayed in Figure \ref{limb} at the $\gamma$-ray energy of 100~MeV. The spectrum is soft and well fitted by the power law $4.13\times E^{-4.25} \text{cm}^{-2} \text{s}^{-1} \text{MeV}^{-1} \text{sr}^{-1}$ where $E$ denotes the $\gamma$-ray energy in MeV. Above 200~MeV its contribution becomes negligible for our analysis. 
%Below this energy this method provided an adequate template for the limb photons which did not bias the fit.

\section{$\gamma$-ray data selection}
\label{sec:Data_Selection_section}
%We derived the interstellar emission model from a fit of Equation \ref{eqRing} to 
For the LAT data, we have used the P7REP Clean class events from the first 4 years of the mission. The Clean selection has a reduced residual background of misclassified charged particles compared to the Source selection \citep{Ackermann:2012p4094}. We have excised time intervals around bright GRBs and solar flares. This time selection exactly matches that for the 3FGL catalog analysis. We have generated the exposure and PSF maps using the P7REP\_V10 \citep{Bregeon:2013p4093} IRFs (see also Section \ref{sec:fitting_procedure_section} and \ref{sec:GIEM_construction_section}). The event selection excluded photons with zenith angles greater than 100$\degr$ and times when the rocking angle of the spacecraft was greater than 52$\degr$ in order to limit contamination from photons produced in the Earth limb. 
We have binned the LAT photon counts into 14 equal logarithmic intervals from 50~MeV to 50~GeV. Below 50~MeV, the combined effects of the worsening energy resolution and the steep dependence of the effective area on energy as well as the increased Earth limb contamination owing to the increased breadth of the PSF make the study of the diffuse emission more difficult. Above 50~GeV the statistics are too low to accurately separate the numerous emission components, 
%to discriminate between the large number of templates that comprise the model, 
especially in the Galactic plane.

\section{$\gamma$-ray model}
\label{sec:fitting_procedure_section}
Because of the ISM transparency to \g rays and of the uniform CR penetration in the \hi, DNM, and CO-bright gas phases, we have modeled the all-sky LAT photon data as a linear combination of $(l,b)$ maps in Galactic coordinates from the emission components presented above, namely:  \hi column densities, $N_{\text{H}\,\textsc{i}_i}(l,b)$, and W(CO) intensities, $W_{CO_i}(l,b)$, in annuli of different Galactocentric radius $R_i$; the total, dust-inferred, gas column density in the DNM, $N_{\rm H}^{\rm DNM}(l,b)$; dust-inferred corrections to the total \hi column densities,  $N_{{\rm H}\textsc{i}}^{\rm corr}(l,b)$; an inverse Compton intensity, $I_{IC_{p}}(l,b,E)$, predicted in direction and energy ($E$) by GALPROP; an isotropic intensity, $I_{iso}$, for the extragalactic and instrumental backgrounds; the accumulated emissions from the Sun and Moon, $I_{\rm Sun\_Moon}(l,b,E)$; the Earth's limb emission,  $I_{\rm limb}(l,b)$; a set of 3FGL point sources in the ($l_j, b_j)$ directions, and a set of extended sources, $I_{ext_k}(l,b)$.

%In addition to the interstellar emission, the LAT detects $\gamma$ rays from other sources that need to be taken into account in the analysis. We do this by adding dedicated components to account for a residual intensity of the Earth limb ($I_{limb}$, see Section \ref{sec:limb_section}), for point sources and extended $\gamma$-ray sources ($I_{ext}$), and for the emission from the Sun and the Moon (${I}_{sun\_moon}$). Here we are considering extended $\gamma$-ray sources to be small extended sources associated with specific supernova remnants, pulsar wind nebulae and spatially-resolved galaxies. Finally we add a uniform intensity template ($I_{iso}$) to account for unresolved extragalactic $\gamma$-ray sources and CR contamination in the data.
For a given map pixel and energy band we have calculated the predicted number of photon counts, $N_{pred}(l,b,E)$, detected by the LAT as:
%{\footnotesize
%\begin{equation}
%\begin{split}
%  N_{pred}(l,b) &= \iint d\Omega_k \bigg(\sum_{i=rings} \left[q_{HI,i} N_{HI}
%(r_i,l_k,b_k) + q_{CO,i} W_{CO}(r_i,l_k,b_k) \right] \\
%  & \phantom{ {=} \int d\Omega_k \bigg(} + q_{DNM} I_{DNM}(l_k,b_k)  +  q_{IC} I_{IC}(l_k,b_k)  +
%I_{iso} \bigg) \epsilon(l_k,b_k) \, PSF(l,b,l_k,b_k)  \\
%  &+ \sum_{j=sources} F_{j} \,  \epsilon(l_j,b_j) \, PSF(l,b,l_j,b_j)
%\end{split}
%\label{eqRing}
%\end{equation}
%}
%{\footnotesize
%\begin{equation}
%\begin{split}
%%N_{pred}(E,l,b) = \sum_{i=templates} q_{tp_{i}}(E)\widetilde{I}_{tp_{i}}(l,b)  +  N_{IC_{p}}(E)\widetilde{I}_{IC_{p}}(E,l,b)  + C_{iso}(E)\widetilde{I}_{iso} + C_{limb}(E)\widetilde{I}_{limb}(l,b)  + \sum_{j=sources} N_{j}(E)\widetilde{\delta}(l,b,j)
%N_{pred}(E,l,b) = \sum_{i=templates} q_{tp_{i}}(E)\widetilde{I}_{tp_{i}}(l,b)  +  C_{IC}(E)\widetilde{I}_{IC_{p}}(E,l,b)  + C_{iso}(E)\widetilde{I}_{iso}   + 
%\sum_{i=ext~src} C_{ext_{i}}(E)\widetilde{I}_{src_{i}}(l,b) \\ + \sum_{i=point~src} N_{pt_{i}}(E)\widetilde{\delta}(l,b,i) 
%+ \widetilde{I}_{sun}(E,l,b) +  \widetilde{I}_{moon}(E,l,b) + C_{limb}(E)\widetilde{I}_{limb}(l,b)
%\end{split}
%\label{eqRing}
%\end{equation}
%}

%      I(l,b,E) &=& q_{\rm LIS}(E) \times \bigr[\, \sum_{i=1}^3 q_{\ion{H}{i},i}(E) \, N_{\ion{H}{i},i}(l,b) \nonumber\\
%      &+&\, q_{\rm{CO}}(E) \, W_{\rm{CO}}(l,b) + \, q_{\rm{DNM}}(E) \, D^{\rm{DNM}}(l,b)\,\bigr] \nonumber \\

{\footnotesize
\begin{equation}
\begin{split}
N_{pred}(l,b,E) &= \sum_{i=annulus} q_{{\rm H}\textsc{i}_i}(E) \widetilde{N}_{{\rm H}\textsc{i}_i}(l,b) + \sum_{i=annulus} q_{{\rm CO}_i}(E) \widetilde{W}_{{\rm CO}_i}(l,b) \\ 
& + q_{\rm DNM}(E) \widetilde{N}_{\rm H}^{\rm DNM}(l,b)  + q_{{\rm H}\textsc{i}\,corr}(E) \widetilde{N}_{{\rm H}\textsc{i}}^{\rm corr}(l,b)  \\
& +  C_{IC_{p}}(E)\widetilde{I}_{IC_{p}}(l,b,E)  +  C_{iso}(E)\widetilde{I}_{iso}  \\
& + C_{\rm limb}(E)\widetilde{I}_{\rm limb}(l,b)  + \widetilde{I}_{\rm Sun\_Moon}(l,b,E) \\ 
& + \sum_{j=point~src} N_{pt_j}(E)\widetilde{\delta}(l-l_j,b-b_j)   +  \sum_{k=extend~src} C_{ext_k}(E)\widetilde{I}_{ext_k}(l,b) 
% + \sum_{i=patch} C_{patch_i}(E)\widetilde{I}_{patch_i}(l,b)       
\end{split}
\label{eqRing}
\end{equation}
} 
% & +  C_{IC_{p}}(E)\widetilde{I}_{IC_{p}}(l,b,E) + C_{\rm NPS}(E)\widetilde{I}_{\rm NPS}(l,b)  +  C_{iso}(E)\widetilde{I}_{iso}  \\

% \sum_{i=patches} N_{patches_{i}}(E)\widetilde{I}_{patches_{i}}(l,b)
%N_{pred}(E,l,b) &= \sum_{i=H~templates} \widetilde{q_{i}(E) N_{H_i}(l,b)}  +  C_{IC_{p}}(E)\widetilde{I}_{IC_{p}}(E,l,b) +  C_{iso}(E)\widetilde{I}_{iso}  \\ &    + C_{LoopI}(E)\widetilde{I}_{LoopI}(l,b) + \sum_{i=patch} C_{patch_i}(E)\widetilde{I}_{patch_i}(l,b)        + C_{limb}(E)\widetilde{I}_{limb}(l,b)  \\ & + \sum_{i=point~src} N_{pt_{i}}(E)\widetilde{\delta}(l,b,i)   +  \sum_{i=extend~src} C_{ext_{i}}(E)\widetilde{I}_{ext_i}(l,b) +  \widetilde{I}_{sun\_moon}(E,l,b) 

%where E is the energy and $l$,$b$ are Galactic sky coordinates. 
In Equation \ref{eqRing}, the $q$ parameters represent the $\gamma$-ray emissivity of the hydrogen in the associated column-density maps and in the CO-bright phase for a given \xco conversion factor. Without distance information, the $q_{\rm DNM}$ and $q_{{\rm H}\textsc{i}\,corr}$ parameters correspond to the assumption of uniform CR densities in the whole DNM and for all \nhi corrections. The $C_{IC_{p}}$, $C_{\rm limb}$, $C_{iso}$, and $C_{ext_k}$ terms represent normalization factors for the associated intensity maps. $N_{pt_{j}}$ denotes the total number of emitted photons per each energy band for each point source
%Equation \ref{eqRing} also incorporates coefficients associated with extended sources () and to point sources ($N_{pt_{i}}$) which are 
represented by the Dirac $\delta$ function. All the $q$, $C$, and $N_{pt_{j}}$ factors are to be determined by fits to the LAT data in each energy band. The free $I_{IC_{p}}$ normalization partially allows for possible variations of the CR and ISRF spatial and energy distributions from what was assumed. The solar and lunar intensities were not allowed to vary in the fits.
 
%To account for unmodeled excesses we also introduced in Equation \ref{eqRing} patches of uniform intensity ($I_{patch_i}$, see Section \ref{sec:Large_scale_structures}) with corresponding normalization factors $C_{patch_i}$. 

We use the tilde notation $\widetilde{I}$ to denote count maps resulting from the convolution with the LAT PSF of the product of an intensity map $I$ and of the instrument exposure and pixel solid angle. We have used the Science Tools
%\footnote{The Science Tools, IRFs, and LAT $\gamma$-ray data are available from \url{http://fermi.gsfc.nasa.gov/ssc/}} 
{\it gtpsf}  and {\it gtexpcube2} to estimate the PSF and the binned exposure with the preliminary set of IRFs P7REP\_CLEAN\_V10. The final model was ultimately scaled to the publicly available P7REP\_CLEAN\_V15, see Section \ref{sec:Resulting_model_section}. Since both exposure and PSF are energy dependent, we have applied Equation \ref{eqRing} in energy bins in which we have averaged the PSF and exposure with a weight corresponding to a power-law spectrum of index 2. This choice has very little impact because the energy bands are narrow.

In order to fit the model to the LAT data, we have used a binned maximum likelihood with Poisson statistics. All maps were binned in HEALPix\footnote{\url{http://healpix.sourceforge.net}} with an $N_{\rm side}$ value of 256, so the bin size is about $0\fdg25 \times 0\fdg25$. We have used the code MINUIT\footnote{\url{https://wwwasdoc.web.cern.ch/wwwasdoc/minuit/minmain.html}} to maximize the logarithm of the likelihood and to calculate the uncertainties on the parameters. 

%The likelihood $L$ is calculated as the product, for all the pixels of the region of interest, of the Poisson probabilities $P_i$ of observing $n_{i}$ photons in a pixel $i$ where the model predicts $\theta_{i}$: 
%{\footnotesize
%\begin{equation}
%\begin{split}
 % L&=\prod\nolimits_{i=pixels} P_i \\
%P_i&=\theta_{i}^{n_{i}}e^{-\theta_{i}}/n_{i}!
%\end{split}
%\label{eqL}
%\end{equation}
%}

We did not fit Equation \ref{eqRing} to the entire all-sky LAT data set at once. We have applied latitude and longitude cuts to define sub-regions where some templates are prominent.
This allows to account for the increasing level of degeneracy between components with decreasing latitude, to optimize the derivation of local versus distant emissivities, and to separate the contributions of structured (gas) versus smooth (e.g. $IC_{p}$, $I_{iso}$) components at the largest angular scales. We describe them with the results in the following sections.

The model includes a comprehensive list of diffuse emission components and of known localized sources, but earlier GIEM versions and the results of a preliminary fit with all components left free have revealed extensive regions of highly significant emissions in excess of the model. Some excesses exhibit patterns relating them to well-known objects such as the \fb and \loopI (along the North Polar Spur, but also in lower-latitude parts of the old supernova remnant). Other bright excesses of unknown origin extend along the Galactic plane, in particular in the first Galactic quadrant at $10\degr \leq l \leq 50\degr$, in the fourth quadrant around $l=315\degr$, and in the Cygnus region. All these excesses are due to the lack of suitable templates in the model and are further discussed in Section \ref{extra_emission}. Hereafter, we generically refer to them as regions of Extended Excess Emissions (EEE). To lower the impact of their existence on the derivation of the modeled components, we have developed specific strategies based either on the addition of uniform patches in the model or on the iterative insertion of residual maps, filtered to remove small angular structures. Because they are statistically highly significant, it is necessary to delineate and account for these diffuse excesses to avoid biasing the spectra of the other components in the model, which would otherwise somewhat compensate for the missing features. We have preferred this approach to masking out large excess zones which could jeopardize the derivation of the other parameters.

\section{Gas emissivities} 
\label{sec:fit_and_interpretation}
\subsection{Beyond 7 kpc in Galactocentric radius} 
\label{sec:local_annuli}
In order to measure the hydrogen emissivity spectra in the various gas phases near the Solar circle and in the outer Galaxy, we have performed a series of successive fits. 
%targetting certain emissivities to be estimated in a specific region, then to be maintained fixed in subsequent fits while all the other model parameters are free to vary. 
We took advantage of the broad extent in latitude of the local gas to reduce the influence of the much brighter Galactic ridge. 

In addition to the NPS radio template, we have built simple intensity patches to account for the sources of EEE. The patches encompass regions with an excess of photons of at least 20\% compared to the best-fit preliminary model (with all parameters in equation \ref{eqRing} left free). This cut is well above the average level of positive or negative residuals found in the rest of the sky.  Figure \ref{fig_patches} shows the location and extent of the seven patches. The first four uniform patches fill the region toward the northern part of \loopI, a disk-shaped one covers the region around the cocoon of hard $\gamma$ rays observed by \cite{Ackermann:2011p2959} toward Cygnus; the last two cover the \fb. Each patch has a spatially uniform intensity spectrum included, with the NPS template, in the fits to the data as additional free components in Equation \ref{eqRing}.
%We added 4 patches spatially associated with \loopI including a large rounded shape filling the loop and three smaller patches closer to the Galactic plane. Additionally, we have created two patches for the FBs. We also made a . As for other templates we assumed no particular spectrum for the patches, but derived them from the fits.
%Table~\ref{tbl:patches} summarizes the patches locations. 
We note that the GIEM does not include the NPS map or the uniform templates.

The results on the gas emissivity spectra for Galactocentric radii greater than 7 kpc have been obtained by applying the following longitude and latitude cuts and by performing the series of successive fits in each of the 14 energy bins in the following order:
\begin{itemize}
	\item \hi emissivity in the 8--10 kpc annulus about the Solar circle, $q_{{\rm H}\textsc{i}_7}(E)$: all longitudes and $10\degr < |b| < 70\degr$
	\item CO emissivities about the Solar circle, $q_{{\rm CO}_{6+7}}(E)$: all longitudes, $4\degr < |b| < 70\degr$, where CO emission is significantly detected in the moment-masked filtered maps. %The broad excess in the CO annulus 6 at longitudes around $40\degr$ coincide with a region of strong EEE. 
	\item gas emissivity in the DNM, $q_{\rm DNM}(E) $, \hi emissivity in the 7--8~kpc annulus, $q_{{\rm H}\textsc{i}_6}(E)$, and emissivity associated with the \nhi corrections: all longitudes and $3\degr < |b| < 70\degr$
	\item \hi and CO emissivities in the outer Galaxy, $q_{{\rm H}\textsc{i}_8}(E)$, $q_{{\rm H}\textsc{i}_9}(E)$, and $q_{{\rm CO}_{8+9}}(E)$: all latitudes and $90\degr < l < 270\degr$.
\end{itemize}
We have checked that the measured emissivity spectra do not significantly depend on the precise shapes of the patches as long as they approximately follow the edges of the EEE and contain most of its emission. The emissivity spectra in the outer annuli are rather insensitive to the patches as the latter gather in the first and fourth Galactic quadrants. Absolute latitudes have several times been restricted to $70\degr$ because the isotropic emission dominates at higher latitudes as described below. For those independent fits we left all the template normalization coefficients of Equation \ref{eqRing} free to vary in each of the 14 energy bins except for the Sun and the Moon templates. 

\subsection{In the inner Galaxy} 
\label{sec:inner_Gal_section}
The inner Galactic region is particularly difficult to model. The gas column densities are most strongly affected by optical depth corrections and self-absorption, and by limited kinetic distance resolution at low longitudes. The determination of dust reddening for the DNM is less precise \citep{Abergel:2013p4126} and we cannot trace CR density variations with distance in this component. Additionally, $\gamma$-ray point-like and extended sources are numerous and the IC$_{p}$ morphology is uncertain. For one or several of these limitations in the interstellar modelling, or because of an over-density of CRs in certain regions, we observe EEE at low latitudes toward
%Possibly for one of those reasons, or because of an excess of CR or an incorrect modeling of a foreground emission, we observed a broad unmodeled emission (referred to as ``extra emission'' in the rest of the text) in the direction of 
the inner Galaxy, with a maximum in the first Galactic quadrant near the base of the North Polar Spur ($l \sim$30$\degr$). To ensure that the EEE are not taken up at low energies by undue softening of the individual point sources because of the broad PSF we have fixed the source intensities to those found in the first iteration of the 3FGL catalog, which was calculated with all components of the diffuse model and patches set free. The use of uniform intensity patches to account for the EEE would lead to a strong dependence of the inner-annuli emissivities on the patch shapes. We have used instead a two-step procedure in which the shape and spectrum of the EEE are iteratively determined from residual emission and incorporated in the $\gamma$-ray model in order to measure the gas emissivities.
%Up to this stage this emission was approximately accounted for by patches of uniform intensity. At this point we removed the patches in Equation \ref{eqRing} and we modeled the extra emission with a two-step iterative procedure.
The $N_{\text{H}\textsc{~i}}$ and $W$(CO) annuli in the inner Galaxy have limited morphological differences. With the added presence of the bright, unknown EEE, it became necessary to reduce the number of free components in the inner-Galaxy parts of the model and to proceed with a single hydrogen template for each annulus.

In the first step, we have taken advantage of the smaller extent and reduced intensity of the EEE in the fourth Galactic quadrant to measure the $X_{\text{CO}}=N_{\text{H}_{2}}$/$W$(CO) conversion factors for the inner annuli 1 to 5, assuming that the $\gamma$-ray emissivity of the H$_{2}$ molecule is twice that of \hi. Under this assumption the \xco factor is half the ratio between the \hi and CO emissivities: $X_{{\rm CO}_i}= q_{{\rm CO}_i}/(2q_{{\rm H}\textsc{i}_i})$. We have fitted Equation \ref{eqRing} to the LAT observations (without patches) in the fourth Galactic quadrant and GC region ($270\degr<l<365\degr$). We have left all the components free to vary, including the local annulus, which could be constrained by the latitude extent, $\left| b \right| <20\degr$, of this fit. This was done to optimize the determination of the gas emissivities in the inner annuli.  
%We obtained a residual map in counts with some emission not accounted for by templates of  Equation \ref{eqRing}.
To account for the EEE, we have selected the positive residuals, we smoothed them with a 2-dimensional Gaussian symmetric kernel of 3$\degr$ FWHM to weaken the correlation with the gas distributions, and we have re-injected them as an additional component, with a free intensity for the next iteration. We have iterated three times up to the point where the distributions of positive and negative residual intensities became comparable. We have selected the \xco values obtained for the energy band centered on 2~GeV where high statistics and the narrow PSF reduce the cross-correlation between the \hi and CO maps. We have obtained  $X_{\text{CO}} = 0.5\times$10$^{20}$~cm$^{-2}$~(K~km~s$^{-1}$)$^{-1}$ for annuli 1 and 2, and $X_{\text{CO}} = 1.5\times$10$^{20}$~cm$^{-2}$~(K~km~s$^{-1}$)$^{-1}$ for annuli 3, 4, and 5. We did not detect a significant signal from the atomic hydrogen in the CMZ, but the gas of this region is largely molecular.

For the second step, we have fitted the whole Galactic disc, $0\degr \leq l < 360\degr$ and $|b| <20\degr$, with the combined $N_{\text{H}}=N_{\text{H}\textsc{~i}}$+2$X_{\text{CO}}$W(CO) maps with free emissivities for the inner annuli 1 to 5, with the W(CO) CMZ map with a free emissivity, with the \hi and CO annuli (not combined) beyond 7 kpc with free emissivities, and with all the other model components of equation \ref{eqRing} left free to vary, except for the sources. We have used the same iterative procedure to account for the EEE. With this fit, we have obtained the \g-ray emissivity spectra per hydrogen atom in the inner annuli. 
%We added to Equation \ref{eqRing} the template for the extra emission derived in step one. As mentioned above, at this stage the patches were removed from Equation \ref{eqRing}. To ensure that the extra emission at low energies is not taken up by fitting the individual point sources we fixed their intensities to those found in the first iteration of the 3FGL catalog calculated with a preliminary interstellar diffuse emission model that included patches. We repeated the iterative procedure described in the first step to derive the final emissivities per hydrogen atom in the inner annuli. We also obtained a template corresponding to the extra emission and large-scale structures.

\subsection{Gas emissivity spectra across the Galaxy}
\label{sec:Gas_emissivities_section}
\begin{figure}[h!]
\begin{center}
\includegraphics[width=17cm]{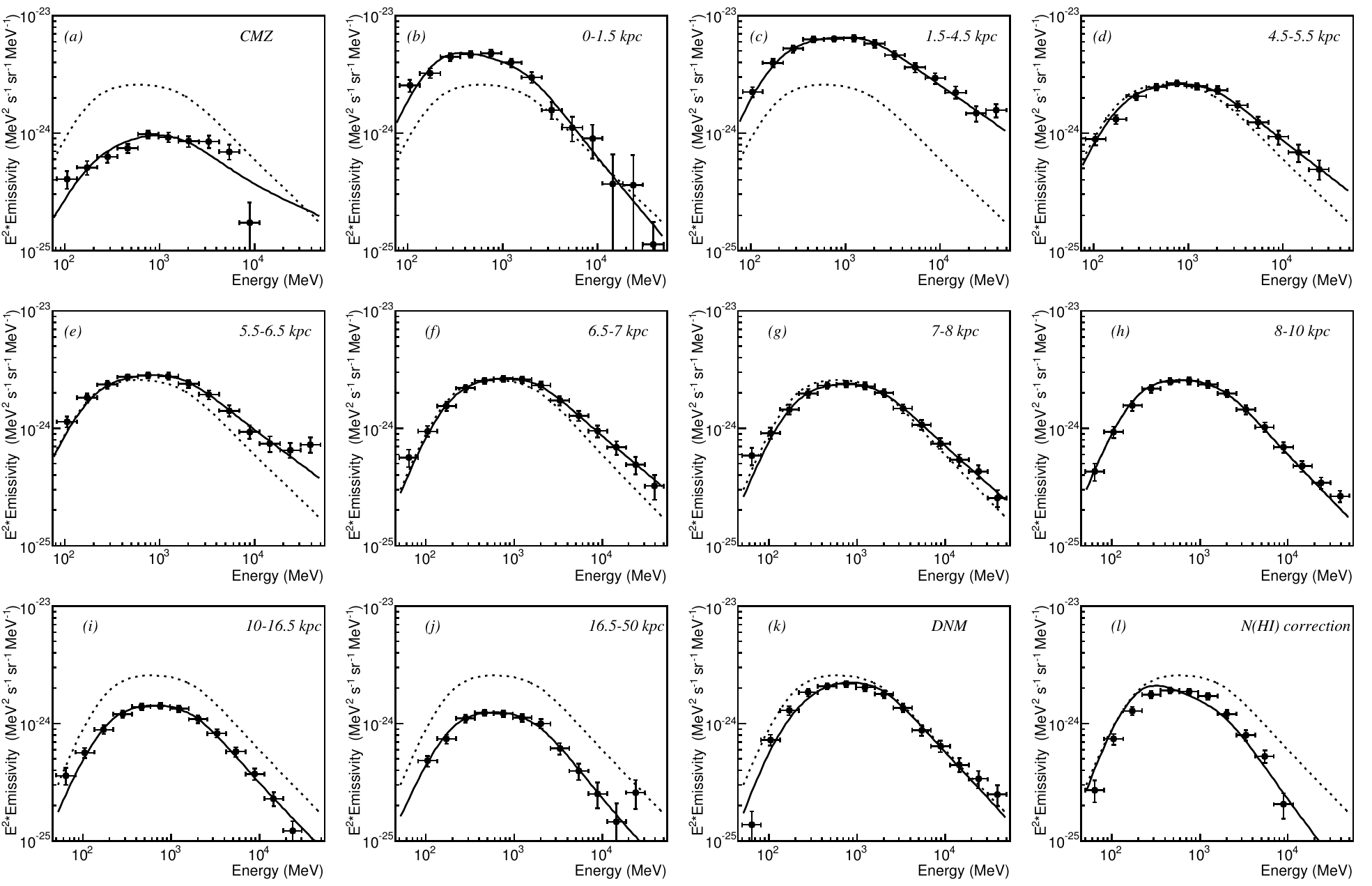}
\caption{(a)-(j): Spectral energy distributions of the $\gamma$-ray emissivity per H atom in the \hi and H$_2$ phases for the CMZ and the nine Galactocentric annuli. The solid curve shows the best fit obtained with a combination of pion emission from CR nuclei and bremsstrahlung radiation from CR electrons. The dashed curve shows the best fit for the local annulus.
% a fitted model based on proton density and production cross-section. 
To display the gas SED in the DNM (k) and that associated with the \nhi correction map (l), we have used a gas-to-dust reddening ratio of 3.5$\times$10$^{21}$ cm$^{-2}$ mag$^{-1}$.
We did not display emissivities below 10$^{-25}$~MeV$^{2}$~s$^{-1}$~sr$^{-1}$~MeV$^{-1}$ or the values for the lowest energy bin for the inner Galaxy annuli. Those points were not used in the analysis.}
\label{gas_emiss}
\end{center}
\end{figure}

Figure \ref{gas_emiss} shows the spectral energy distributions (SEDs) of emissivities obtained for the nine Galactocentric annuli and for the CMZ region from the differential $\gamma$-ray emissivities per hydrogen atom $\frac{dq}{dE}=q/\Delta E$, where $\Delta E$ is the energy bin width. We did not correct the emissivities for the LAT energy dispersion. The emissivities plotted in Figure \ref{gas_emiss} overestimate the true values by approximately 10\% at 100~MeV and 50\% at 50~MeV, it is negligible above 200~MeV \citep{Casandjian:2015p4380}. As described above, the emissivity per hydrogen atom is derived in the \hi phase for annuli 6 to 9 and in both the atomic and molecular hydrogen phases for annuli 1 to 5. The emissivity in the CMZ, measured from the CO map, was scaled to emissivity per hydrogen atom assuming $X_{\text{CO}}=0.5\times$10$^{20}$~cm$^{-2}$~(K~km~s$^{-1}$)$^{-1}$, the same as what we found for annuli 1 and 2. The low \xco value of $0.2-0.7\times$10$^{20}$~cm$^{-2}$~(K~km~s$^{-1}$)$^{-1}$ inferred from infrared observations by \cite{Sodroski:1995p4124} in the GC region also supports this choice. We observe that the emissivities follow continuous energy distributions even though the $\gamma$-ray fits were performed independently in each energy bin. The error bars plotted in Figure \ref{gas_emiss} represent only the statistical uncertainties. 

For the local ISM, we show in \cite{Casandjian:2015p4380} that the systematic uncertainties associated with the measurement of the hydrogen $\gamma$-ray emissivity are dominated by the uncertainty in the LAT effective area over the whole energy range. It amounts to 10\% below 100~MeV, with a linear decrease in $\log(E)$ to 5\% in the range between 316~MeV and 10~GeV, and a linear increase in $\log(E)$ up to 15\% at 1~TeV\footnote{\url{http://fermi.gsfc.nasa.gov/ssc/data/analysis/LAT\_caveats.html}}. Those values represent a lower limit for the hydrogen emissivities in the outer Galaxy where the non-uniformity of $T_{S}$ cannot be neglected. 

For the hydrogen emissivities in the inner annuli ($<7$~kpc), the major source of systematic uncertainties is the model incompleteness. We note that when we restrict the fit to the fourth Galactic quadrant without using any residual template or patch for the EEE, the gas emissivities in the CMZ and in inner annuli 2 and 4 increase globally by up to 40\% compared to those given in Figure \ref{gas_emiss}. A second source of uncertainty is the amount of dense, cold \hi in the inner spiral arms and the spin temperature correction. The \hi mean opacity increases inward from the solar circle and peaks in the molecular ring, suggesting that the cold phase is more abundant and colder there than it is locally \citep{Dickey:2003p801}. However, the median fraction of cold \hi is about 20\% in column density \citep{Heiles:2003p1829,Murray:2015p89}, so the measurements of the \g-ray emissivities per H atom are constrained primarily by the gas content in the warm \hi phase which is more reliably quantified. A change in $T_S$ from 140~K to 400~K \citep{Dickey:2009p1250} results in a $\sim$30\% change in \nhi for the inner annuli. A third source of uncertainty is the quality of the dust map at low latitudes which is hampered by the poor angular resolution of the temperature corrections against a rising spatial density of warm star-forming regions. There exist localized differences by a factor up to 2 or 3 along the Galactic plane between the estimates recently inferred at 5$\arcmin$ with Planck \citep{Abergel:2013p4126} and the \cite{Schlegel:1998p290} map available at the start of this work; the outer Galaxy reddening is also significantly underestimated in the \cite{Schlegel:1998p290} map while the inner Galaxy is too dusty. These differences do not affect so much the gas emissivity in the DNM (because of the presence of massive clouds off the plane to help the fits) as the \nhi correction. We estimate a 30--40\% uncertainty which can partially propagate to the uncertainties of the emissivities in the inner annuli.
%That includes the emission of the missing DNM and the extra emission that peaks at a longitude of $\sim$30$\degr$. As we saw in Section \ref{sec:inner_Gal_section} we did not use any patch to derive the inner annuli emissivities, that would have lead to a too strong dependence on the patches shape. Instead we used a two-step procedure in which the shape and spectrum of the unmodeled emission was first determined iteratively and then this emission was incorporated in the $\gamma$-ray fit to get the gas emissivities. The systematic uncertainties of the inner annuli emissivity depends on this procedure, as well as 
%\isa{HOW MUCH UNCERTAINTY FROM the correlation of the gas annuli with $I_{IC_{p}}$ and with the point and extended sources ??? 
%We therefore expect the systematic uncertainties on the emissivities of the inner annuli to be of order XXX \% at GeV energies. MUCH MORE xxx AT LOW ENERGY.}
%\isa{
%I DON'T KNOW THE LEVEL OF CORRELATION BETWEEN THE INNER EMISSIVITY AND THE SRC AND THE IC, SEEMS TO ME LIKE A FAIRLY BIG WORK TO DO
%} 

%To interpolate the emissivities between the measured points and extrapolate above 50~GeV 
We have fitted the differential emissivity SEDs with a model of bremsstrahlung emission \citep{Gould:1969p2474} and hadronic decay. Most $\gamma$ rays with energies between 100~MeV and 50~GeV originate from the decay of $\pi^{0}$ produced in hadronic collisions when CR protons with energies above 0.5~GeV interact with protons from ISM nuclei \citep{Stecker:1970p3081, AguilarBenitez:1991p4004}. We have fitted the emissivity SED of each annulus between 200~MeV and 30~GeV using the $\gamma$-ray production cross-section of \cite{Kamae:2006p2590} and a CR proton flux spectrum parametrized as: $A\beta^{P_1}R^{P_2}$ where $\beta=v/c$ is the proton-to-light velocity ratio, $R$ is the proton rigidity, $A$ is a free normalization, and $P_1$ and $P_2$ are free spectral indices \citep{Shikaze:2007p3492}. This form tends to a power law with index $P_2$ at high rigidity (energy) and the $P_1$ index controls the spectral fall-off at low rigidities which is indicated by the Voyager 1 data near the heliopause and by \g-ray measurements in the local ISM \citep{Grenier:2015p4403,Casandjian:2015p4380}. We have used the results of \cite{Mori:2009p341} to scale the proton-proton cross-section to the nucleus-nucleus one, taking into account the abundance of heavier nuclei in the ISM and in the CRs \cite{Casandjian:2015p4380}. We have accounted for the bremsstrahlung radiation contribution using the following electron spectral form: $B(E_{kin}/E_0+((E_{kin}-E_4)/E_0)^{-0.5})^{P_3}$ where $E_{kin}$ is the kinetic energy of the electrons, $E_0=1$~GeV, $B$ is a free normalization, $P_3$ is a free spectral index, and $E_4$ is a free scaling energy. We have fitted the bremsstrahlung emission together with the hadron decay component to the emissivity spectrum measured in the local annulus 7 (8--10~kpc). We have found that the modeled emissivities systematically underestimate the measured ones by about 20\% above 2 GeV. Yet, the measured emissivity spectrum compares well, within $\sim$10\%, with previous measurements obtained with the same $T_S$ in specific regions of the local ISM where there is minimal confusion along the lines of sight and higher resolution ISM data were used \citep[see Figure 4 of][]{Grenier:2015p4403}. 
For every annuli we have corrected for this discrepancy by increasing the Kamae et al. production cross-section by 20\% for $\gamma$-ray energies above 2~GeV, with a smooth transition to lower energies to avoid any discontinuity. Even though we have applied this correction to the proton-proton cross-section, it could originate from an incorrect scaling factor from the proton-proton to nucleus-nucleus cross section, or from the parametrization of the proton spectrum. A detailed interpretation of the emissivity SED, compatible with the one presented here, derived in the local annulus from a similar template fitting method is given in \cite{Casandjian:2015p4380}.
%The origin of the correction does not affect the final model. 

From the CR spectra fit to the emissivity SED in the local annulus, we have derived the bremsstrahlung contribution, and the proton functional parameter $P_1$ which describes the low-rigidity turn-over of the proton spectrum. We have assigned those same values to the other annuli. The bremsstrahlung radiation contributes only 14\% at 200 MeV to the total emissivity spectrum in the local ISM \citep{Casandjian:2015p4380}. We have interpreted the pion-decay emission above 1 GeV to study potential variations in flux and spectrum of the bulk of the CRs pervading the Galaxy. The corresponding CRs have energies per nucleon well above the low-energy turn-over  \citep[see Figure 1 of][]{Grenier:2015p4403}. At GeV energies and above, the improved LAT performance enables a better separation of the emissions originating from the different annuli. 
%While there is evidence in the radio synchrotron maps for xxx times larger CR electron fluxes in the CMZ (REF) and xxx times larger in the inner Galaxy (REF) than locally, we have no information on the associated proton flux to claim variations in the e/p ratio.
We have fitted the measured SEDs above 280~MeV with the assumption of a uniform CR electron flux and $P_1$ across the Galaxy. As shown in Figure \ref{gas_emiss} we obtained reasonable fit to the SEDs. The derived values of the proton index $P_2$ and normalization $A$ are fairly insensitive to the approximations on $P_1$ and the bremsstrahlung contributions. We tested these approximations by also fitting the SEDs above 100~MeV with the electron normalization and index parameters free and found no significant variation in $P_2$ and $A$.
%\isa{This approximation is justified here because ... ... While there is evidence in the radio synchrotron maps for xxx times larger CR electron fluxes in the CMZ (REF) and xxx times larger in the inner Galaxy (REF) than locally, we have no information on the associated proton flux to claim variations in the e/p ratio (what about GALPROP and DRAGON predictions ???).... JE N'AI PAS UTILISÉ CES PHRASES}
%We then fitted annuli 1 to 6, 8, 9 and the CMZ with only two parameters: the proton spectrum normalization $A$ and proton spectral index $P_2$. To simplify the fit we did not allow any radial variation for the electron density. 
%We observe a good agreement with the experimental emissivities despite the approximations used in the fitting procedure. 
In order to compare in Figure \ref{gas_emiss} the gas emissivity spectrum in the DNM and in the local annulus, we have used a gas-to-dust reddening ratio of 3.5$\times$10$^{21}$ cm$^{-2}$ mag$^{-1}$ \citep{Grenier:2005p836}.
%\isa{we have used the standard gas-to-reddening ratio $N_{{\rm H}}/E(B-V)$ of \cite{Bohlin:1978p2105} (5.8e21 POUR TES FIGURES).  We observe that the SED compares very well with that of the hydrogen in the local ISM, thus providing further evidence for the gaseous origin of this component. It is xxx times brighter/fainter than or fully compatible with the local spectrum $+$ comments... XXX EN FAIT DIFF DE ENVIRON 25\% CAR CARTE EBV UN PEU DIFFERENTES (J'AI FAIT UN SEUIL POUR ENLEVER LOW MAG DANS L UNE DES DEUX)} 
%We applied the same method to fit the coefficients associated with the DNM and $N_{\text{H}\textsc{~i}}$ correction templates obtained from the positive and negative dust residuals. 
In the case of the \nhi correction template, we have used the same $N_{{\rm H}}/E(B-V)$ ratio and we have left all the spectral parameters for electrons and protons free to obtain a better fit to the data. 
%\isa{ADD DISCUSSION ON THE AWKWARD SPECTRUM OF NHICORR AT THE LOWEST AND HIGHEST ENERGIES...} 
%\red{LE PROGRAMME DE FIT TOUT LIBRE, ELECTRONS ET PROTONS, EXISTE ! CE SERAIT VRAIMENT IMPORTANT DE REFAIRE LES FITS DE LA FIGURE \ref{gas_emiss} POUR FORTEMENT CREDIBILISER TOUTE LA PARTIE PHYSIQUE DU PAPIER, SUR LES SPECTRES ET GRADIENTS. COURAGE !!! CE N'EST PAS LOURD PUISQUE LE PROGRAMME EXISTE. ON N'AURAIT PLUS QUE LES FITS 'BLOQUES" POUR L'EXTRAPOLATION POUR LE GIEM ET LA JUSTIFICATION EST ALORS SIMPLE: DU MOMENT QUE CELA FIT BIEN, C'EST VALABLE. J'INSITE CAR JE PARIE QUE LES REFEREES VONT RECLAMER DE REFAIRE CETTE PARTIE. AUTANT ANTICIPER, AVOIR LA PAIX ET FAIRE ACCEPTER LE PAPIER PLUS VITE AINSI. ENCORE UN PETIT EFFORT POUR LA POSTERITE !}
%Figure \ref{gas_emiss} (k) and (l) show the emissivities associated with the DNM and the $N_{\text{H}\textsc{~i}}$ correction templates together with the fit. To display in this graph the emissivities inferred from the dust optical de	pth map in the same units as those from the column-density maps, we divided the measured and fitted emissivities by an nominal gas-to-dust ratio of $4\times10^{21}$ cm$^{-2}$ mag$^{-1}$.

\subsection{Gradients of cosmic-ray spectra across the Galaxy}
\label{sec:gradients_across_the_Galaxy}

\begin{figure}[h]
\begin{center}
\includegraphics[width=18cm]{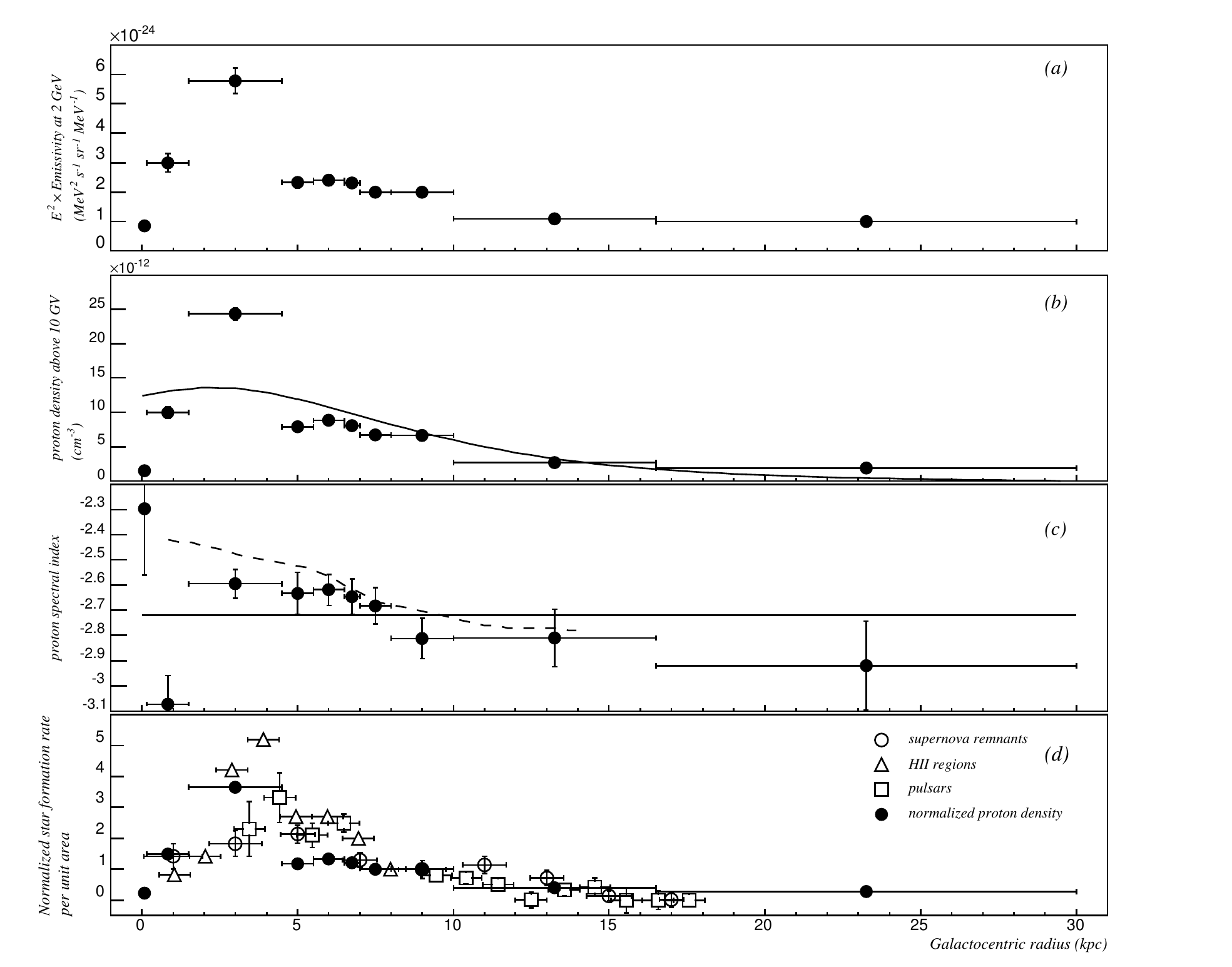}
\caption{Radial distributions across the Galaxy of (a) the \g-ray emissivity per H atom measured at 2 GeV; (b) the proton flux integrated above 10~GV, with the prediction from the GALPROP model $^SY^Z6^R30^T150^C2$ \citep[solid curve,][]{Ackermann:2012p2978}; (c) the proton spectral index, $P_2$, with statistical error bars and the prediction for proton rigidities above 1~TV from the same GALPROP model (solid line) and from \cite{Gaggero:2015p4237} (dashed line). In all plots, the horizontal bars span the radial widths of the gas annuli used for the measurements. The two data points with smallest Galactocentric radii have large systematic uncertainties (see text). Panel (d) shows the proton flux integrated above 10~GV, normalized to its value at the Sun Galactocentric radius, with the star formation rate traced  by supernova remnants, \hii regions, and pulsars \citep{Stahler:2005p4407}.}
\label{new_fig_gradients}
\end{center}
\end{figure}

Figure \ref{new_fig_gradients}a shows the radial distributions across the Galaxy of the \g-ray emissivity measured at 2 GeV and Figure \ref{new_fig_gradients}b the radial distribution of proton density integrated above 10~GV. 
%evaluated from the present work. 
We observe a marked increase in CR density around 3~kpc from the GC.
The {\it Fermi}-LAT counts associated to the second gas annulus dominate the region within $\pm 30\degr$ in longitude. The number of counts integrated above 2 GeV and associated to this annulus is twice that of 3FGL sources and 4 times that of IC in a region defined by the annulus contours. The EEE represents a few percent of the total in this region. This increase around 3~kpc might be associated with an enhanced CR production in the molecular ring. The steep increase relative to the next annulus is reminiscent of the marked increase in star-formation rate indicated by massive stars \citep[\hii regions, supernova remnants, pulsars; see][]{Stahler:2005p4407}, as shown in Figure \ref{new_fig_gradients}d. 
%Part of this increase can be due to a contamination from the extra emission, that we estimated via the iterative procedure described above, in annulus number 2 that extends to $\pm$30$\degr$ in longitude. In Figure \ref{fig_gradients} we also show 

%We also observe in Figure \ref{fig_gradients} that the inferred CR proton density in the CMZ is about 4 times lower than the local one (about 8 times lower if we assume the same \xco as for the local annulus). 
%This is similar to what was observed with {\it COS-B} by \cite{Blitz:1985p4121} where a lower \xco was suggested to explain the anomalously low $\gamma$-ray production compared to the CO line intensity. NO !!! It is due to the difference in CO width in the tidally disrupted clouds near the GC !!!
%The inferred proton density in the CMZ should be interpreted with caution given possible confusion with point sources or with the extra emission, especially at low energies.

The proton density profile predicted by the GALPROP model $^SY^Z6^R30^T150^C2$ \citep{Ackermann:2012p2978} reproduces the trend with deviations from measurements by a factor of 2 at the maximum in the molecular ring region. For Galactocentric distances greater than 5~kpc, the predicted proton density gradient is steeper than the observed one. 
%The GALPROP model also has a broader distribution at $\sim$3~kpc from the GC. A flatter radial distribution of CRs in the Galaxy is well-known
This discrepancy, referred to as `the CR gradient problem' has been known since the \g-ray surveys made by {\it COS-B} \citep{Bloemen:1986p3811, Strong:1988p3093} and {\it EGRET} \citep{Strong:1996p1032, Hunter:1997p329}. \cite{Bloemen:1993p4226} suggested that the radial distribution of CR sources may be flatter than inferred from pulsar and supernova-remnant observations, or that the diffusion parameters derived from the local CR measurements are not the same throughout the Galaxy. A solution to this issue in terms of CR-driven Galactic winds and anisotropic diffusion has been proposed by \cite{Breitschwerdt:2002p4227}. \cite{Uhlig:2012p4228} note that CR-driven winds could also suppress the star formation rate by a significant factor. \cite{Shibata:2007p3958} proposed a non-uniform diffusion coefficient that increases with Galactocentric radius and distance from the Galactic plane. Indeed, models with a large Galactic halo and thus faster CR diffusion are able to better reproduce the $\gamma$-ray emissivity in the outer Galaxy \citep{Ackermann:2012p2978}. 
Such a position-dependent CR diffusion coefficient, linked to the ambient power in the magnetic turbulence induced by stellar and supernova activity, allows for a good reproduction of both {\it Fermi}-LAT and local CR observables \cite[][and references therein]{Gaggero:2015p4237}.
%was used by \cite{Evoli:2012p4056} to interpret the \hi emissivity measured by the LAT in the outer Galaxy \citep{Abdo:2010p3307,Ackermann:2011p4125}. 
On the other hand, more molecular gas in the outer Galaxy is still being found \citep{Sun:2015p4229}, implying that the star formation rate and thus the density of CR sources may be underestimated at large distances. An increase in \xco with distance beyond the solar circle is expected because of the metallicity gradient \citep{Pineda:2013p4081}, but it cannot explain the large \g-ray emissivity values found in the outer Galaxy in correlation with the \hi gas \citep{Abdo:2010p3307}, contrary to what has been proposed by \cite{Strong:2004p1030}. The DNM gas at the \hi--H$_2$ interface, however, is more abundant than the CO-bright H$_2$  beyond the solar circle \citep{Abergel:2011p3177,Pineda:2013p4081} and it should better correlate with \hi spatially. It can offer an alternative or complementary solution to the CR gradient problem that can be tested with forthcoming radio line and dust extinction surveys.
We have plotted the proton spectral index $P_2$  versus the Galactocentric radius of the annulus in Figure \ref{new_fig_gradients}c. The index of $2.81 \pm 0.08$ found in the local annulus is fully consistent with the value of $2.820 \pm 0.003$ (stat.) $\pm 0.005$ (syst.) measured by the PAMELA experiment in the range 30 GV to 1.2 TV, beyond the influence of solar modulation \citep{Adriani:2011p4018}. This agreement indicates that the same CR population pervades the local spiral arm and the immediate solar neighbourhood. Measurements of $\gamma$-ray emissivity in nearby clouds had indicated such a spectral uniformity out to a few hundred parsecs from the Sun \citep[for a review see section 4 of ][]{Grenier:2015p4403}. The present result extends this uniformity to the Local Arm, which dominates the local annulus. 
%The local index is also consistent with the GALPROP prediction tuned to match the larger body of spectral information on CR primary and secondary nuclei near or in the Solar System (see Figure \ref{fig_gradients}). 

We observe a hardening of the proton spectra when moving from the outskirts of the Galaxy to the inner molecular ring. This spectral hardening cannot be due to contamination by the EEE at low latitudes since their emission contributes on average only 10\% to the gas emission. Contamination by unresolved sources such as pulsars is also unlikely. They generally have much harder spectra than the interstellar emission \citep{Abdo:2013p4409}, but they contribute at best a few percent of the diffuse emission (see section \ref{section_pt_sources}). One would need a huge increase in source density in the inner Galaxy, at variance with pulsar and supernova-remnant observations. The proton spectral indices extracted in the CMZ and the first annulus (first two radial bins in Figure \ref{new_fig_gradients}) are evaluated for a region extending $\pm$10$\degr$ from the GC which is dominated by the emission of point sources listed in the 3FGL catalog, together with IC; in this region the confusion with gas emission is maximal. Moreover, as we discussed in Section \ref{sec:Atomic_hydrogen}, in this region the gas column density is calculated in part using interpolations of adjacent regions  and may therefore be inaccurate.
%with other templates or point sources is more likely than for the rest of the sky. 
%the fits may still be contaminated by the hard emission from the \fb, not totally suppressed by the iterative fit procedure (see section \ref{extra_emission}). 

CR transport models such as GALPROP do not predict spectral variations across the Galaxy because they assume uniform diffusion properties and a uniform injection spectrum from CR sources. \cite{Gaggero:2015p4237} have also recently noted a gradual CR hardening toward the inner regions under the cruder assumption that the \g radiation originating from the pion decays in the gas dominates all other emission components, so that the diffuse \g-ray spectrum above 5 GeV directly maps the CR spectrum. They propose to explain this hardening by varying the CR diffusion properties through the Galaxy, specifically by linearly decreasing the rigidity index $\delta$ of the diffusion coefficient toward the GC. This decrease allows for harder CR spectra at small Galactic radii and can also explain the emission deficit noted above a few GeV  in the (inner) disc with uniform CR transport models. Their model also includes strong convection within 6.5 kpc from the GC. Energy-independent CR transport by Galactic winds keeps the spectrum near the hard distribution injected by the sources, so an increasing wind toward the inner Galaxy would eventually dominate over energy-dependent diffusion. Convection in the form of Galactic winds is supported by X-ray \citep{Everett:2008p4408} and radio \citep{McClureGriffiths:2013p4439} observations. Figure \ref{new_fig_gradients}c compares the spectral variation across the Galaxy observed by {\it Fermi}-LAT data with the one predicted by \cite{Gaggero:2015p4237}. We observe a reasonable agreement.

%\begin{figure}[h]
%\begin{center}
%\caption{ 
%We did not include any systematic uncertainty in the integrated proton flux error bar. Those systematic uncertainties could be significant for example close to the GC where the integrated proton flux depends on the \xco ratio used in the analysis.}
%\end{center}
%\end{figure}

\section{Normalization of the inverse-Compton radiation}
\label{sec:ICnorm}
\begin{figure}[h]
\begin{center}
\includegraphics[width=12cm]{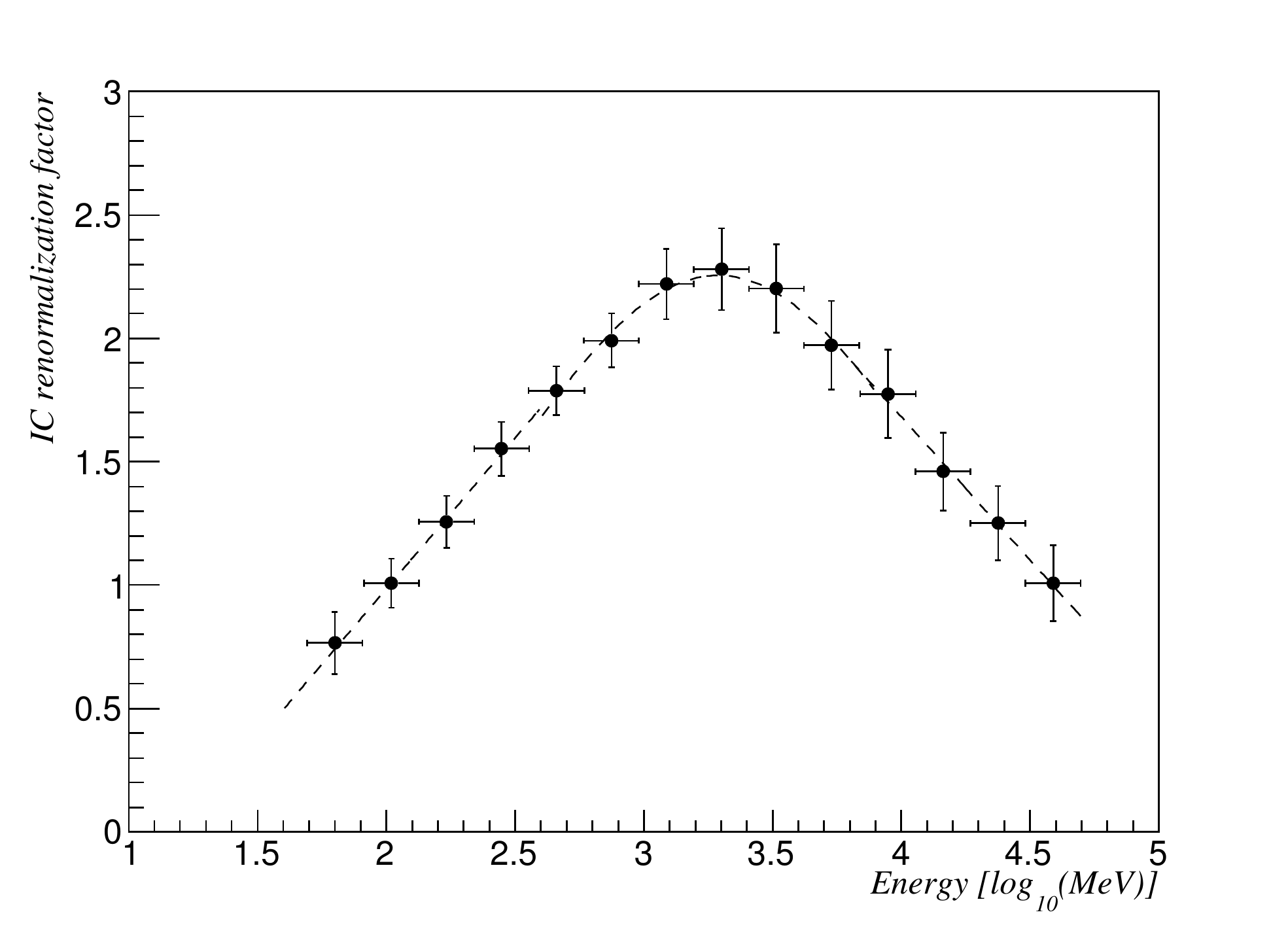}
\caption{Spectral evolution of the normalization factors applied to the GALPROP prediction of the Galactic inverse Compton intensity. The error bars are those obtained from the fit likelihood maximization. The dashed line shows the interpolation used for the construction of the GIEM.}
%between the measured IC normalization coefficients. We used this interpolation to build the IC${_p}$ component for the final interstellar model for energies below 50~GeV. Above this limit we used the nominal GALPROP predictions.}
\label{IC_renorm}
\end{center}
\end{figure}

To study the normalization of the model of the IC intensity, with the spatial and spectral distributions predicted by GALPROP (see Section \ref{sec:Galactic_Inverse_Compton_radiation}), we have fixed the emissivities of the highly-structured components, namely the \hi, CO,  and dust-related, at the values previously measured (see Sections \ref{sec:local_annuli} and \ref{sec:inner_Gal_section}). For each band we have fitted the whole sky with Equation \ref{eqRing}, leaving free all the smooth components with large angular scales, namely the IC, isotropic, and Earth limb emissions, as well as the source fluxes because of their broad effective PSFs at low energies. As before, we have iteratively smoothed the positive residuals in each energy bin and added them to the previous interation residuals until the IC normalization coefficient remained constant. Figure \ref{IC_renorm} shows the normalization factors obtained for the 14 energy bins. The values close to one found near 100~MeV and at the highest energies indicate that no major modification of the GALPROP prediction is required.  
%It steadily increases with  energy up to a maximum of 2.3 at 3~GeV and decreases back to 1 at higher energies. 
The prediction is off by about a factor of two at the intermediate energies. Given the complexity of predicting the leptonic production and propagation as well as the calculation of the ISRF in the Galaxy, the agreement can be considered satisfactory. A more detailed discussion of the comparison between GALPROP IC predictions with various initial conditions is given in \cite{Ackermann:2012p2978}. They also concluded that a greater IC intensity was needed for all models they considered, in particular in the inner Galaxy, either from an increased ISRF, more CR electron sources in the inner regions, or a larger Galactic halo. The present choice of a 6~kpc halo and of a radial distribution for the CR sources strongly peaking near 3 kpc (inside the inner molecular ring) provides larger IC intensities than broader source distributions encompassing the molecular ring \citep{Ackermann:2012p2978}. To investigate the spectral dependence of the normalization factor, one would need to separate the Cosmic Microwave Background (CMB) and dust parts of the IC prediction in Equation \ref{eqRing} to test whether it requires an increase in the very-far-infrared interstellar radiation field with respect to the model of \cite{Porter:2008p3784}. Further investigations on the spectrum and source distribution of the CR electrons should await more precise spatial and spectral models of the \g-ray emissions from \loopI and the \fb.

\section{Regions of EEE}
\label{extra_emission}
\begin{figure}
\begin{center}
\includegraphics[width=13cm]{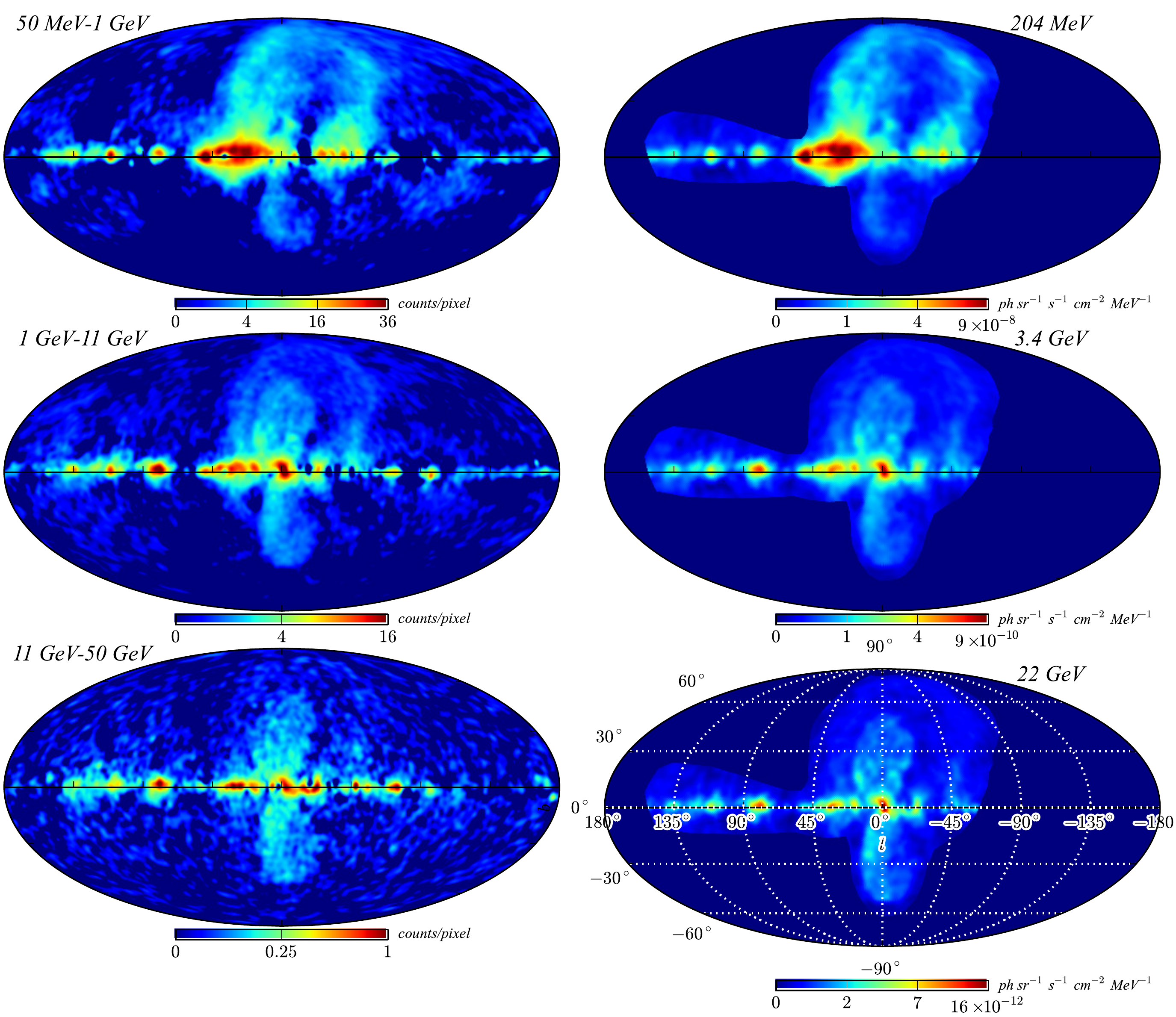}
\caption{\textit{Left column}: Mollweide projection in Galactic coordinates of the EEE found in the energy bands 50~MeV--1~GeV (top), 1--11~GeV (middle), and 11--50~GeV (bottom): the {\it Fermi}-LAT count maps have been obtained after subtraction of the baseline interstellar model described in section \ref{extra_emission} and they have been smoothed with a 2-dimensional symmetric Gaussian of 3$\degr$ FWHM.
%point\isa{-like} and extended sources, the \isa{Earth} limb and isotropic emissions, and a conventional interstellar model based on fitted gas emissivities and scaled IC$_p$ only. 
\textit{Right column}: photon specific intensity, at energies 204~MeV (top), 3.4~GeV (middle), and 22~GeV (bottom), of $I_{EEE}$ that has been developed to describe the EEE at angular scales larger than $2\degr$. All the maps are displayed with a square root scaling and a pixel size of $0\fdg25$.}
\label{fig_comparaison_res_model}
\end{center}
\end{figure}

To display the spatial distributions of the EEE across the sky, we have built a photon count map using Equation \ref{eqRing} with the gas emissivities measured in the different annuli (as in sections \ref{sec:local_annuli} and \ref{sec:inner_Gal_section}), with the IC normalization factors derived in section \ref{sec:ICnorm}, with the isotropic and Earth limb intensities derived from the local annulus fit, with the Sun and Moon intensities, with the fluxes of the point-like and small-extended sources from the preliminary version of the 3FGL catalog. The emission associated with the radio map of the NPS was not included in the calculation since it provides a very limited description of the potential emission originating from \loopI. We refer to this model as the `baseline' one.
%To include in our interstellar emission model the $\gamma$ rays produced by phenomena that lack templates like the large-scale structures, we first created a conventional interstellar emission model based on gas emissivities and IC obtained as described above. We added the sources from the preliminary version of the 3FGL catalog, the predicted Sun and Moon intensities and used an isotropic and normalization of the residual Earth limb component derived from the local annulus fit. 
Figure \ref{fig_comparaison_res_model} (left column) shows the positive difference between the LAT count map and the count map obtained with this model in three energy bands: 50~MeV--1~GeV, 1--11~GeV, and 11--50~GeV. We did not observe strong negative residuals except in the direction of the Carina arm tangent where the model greatly over-predicts the observations.

Figure \ref{fig_comparaison_res_model} exhibits coherent emission features across the sky. They include the NPS and broader \loopI which dominate at medium-to-high latitudes at low energy, and the \fb and strong emission toward their base which are both conspicuous above 1~GeV. Spurs of emission visible at medium northern latitudes toward `the interior' of the NPS (roughly within $-10\degr \leq l \leq 30\degr$ and $10\degr \leq b \leq 30\degr$) spatially relate to a structure in the local DNM gas distribution, thereby indicating the need for more gas than described with the dust reddening used in this work, or a possible enhancement in the CR flux.

We also observe extended sources broadly distributed along the plane at longitudes less than 50$\degr$ and to a lesser extent at longitudes around 315$\degr$. The origin of these excesses is not known. Part of the excess may be caused at low energy by \loopI in the foreground of the Galactic disc.
%It does not seem to be correlated with any gas or dust template and might therefore be of IC origin or from a very nearby gas cloud.  
Strong radio recombination line emission has been detected for longitudes around 30$\degr$ and 330$\degr$ by \cite{Alves:2012p4136} and \cite{Alves:2015p4238}, so the excess of \g radiation could also partly relate to ionized gas.   %This longitude coincides with the peak of the extra emission component that could therefore be associated with H$^{+}$. 
The asymmetry observed between the first and fourth Galactic quadrants below 10 GeV could also have a Galactic IC origin since the GALPROP calculation uses a cylindrical geometry instead of the tilted bar and spiral arms of our Galaxy.
%or be a consequence of the lack of Galactic structure in the CR source distribution used in the GALPROP calculation and therefore in the resulting IC intensity $I_{IC_{p}}$. 
Other extended excesses are present at low latitudes along the Galactic plane. 
%Those seen toward the Cygnus region and near $285\degr$ and $310\degr$ in longitude may be due to an over-correction of the \nhi column densities as inferred from the dust. These excesses will be shortly confirmed or infirmed with the use of more precise dust column densities in new iterations of the LAT GIEM. \isa{non car ils sont la aussi si on ajoute pas la composante negative}

% It is spatially associated with the intersection of \loopI and the Galactic plane. 

\begin{figure}
\begin{center}
\includegraphics[width=13.5cm]{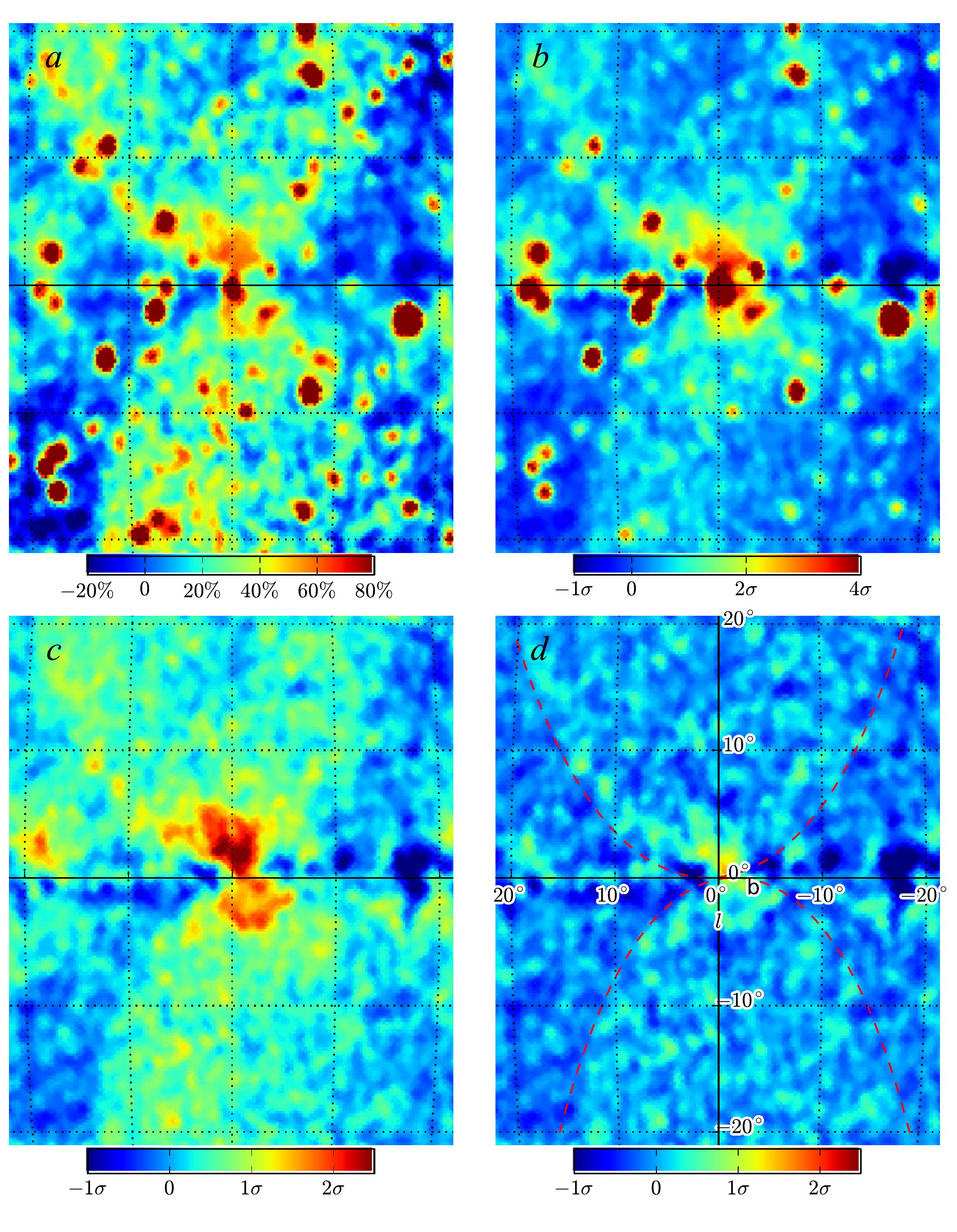}
\caption{Close-up view of a region within 20$\degr$ of the GC showing the {\it Fermi}-LAT count maps integrated between 1.7~GeV and 50~GeV after subtracting the baseline interstellar model described in section \ref{extra_emission}, excluding (top row) and including (bottom row) the point and extended sources from a preliminary 3FGL list in the model. To reduce the emission contrast in latitude, we display the residuals in fractional units (a), dividing the residuals by the model, and in units of standard deviation (b), dividing the residuals by the square root of the model. In (d) we show the residual map after the further subtraction of $I_{EEE}$; it contains structures smaller than the angular scale included $I_{EEE}$. The red dashed lines correspond to the catenary functions that reproduce approximately the edge of the \fb for latitudes below 20$\degr$ (see text for details). We have smoothed the four maps with a Gaussian of 1$\degr$ FWHM.}
\label{gal_center}
\end{center}
\end{figure}

Figure \ref{gal_center} shows a close-up view of the fractional excesses found above the baseline model toward the GC region between 1.7~GeV and 50~GeV. Since the inner Galaxy hosts many point sources, we show the excesses with and without the point-source contribution in the model.
%GC region representing the difference between the LAT counts integrated between 1.7~GeV and 50~GeV toward the GC region and those expected from the conventional model using only the gas and IC components determined in the fits. 
%To reduce the contrast due to the bright $\gamma$-ray emission of the Galactic plane, we divided this difference by the conventional model counts (Figure \ref{gal_center}a) and by the square root of this number (Figure \ref{gal_center}b,c,d). The fluxes of some preliminary 3FGL sources located in the Galactic ridge depend to some extent on the interstellar emission model, so a fraction of the interstellar emission can be incorrectly assigned to point sources. To avoid any bias  we did not subtract them from the counts map in Figure \ref{gal_center}a and \ref{gal_center}b.  They were subtracted in Figure \ref{gal_center}c and \ref{gal_center}d. 
The \fb are clearly visible in Figure \ref{gal_center}a,b,c. We have fit the edges of the bubbles within 20$\degr$ of the GC using various mathematical curves.
The edges of the \fb are well reproduced by two catenary curves: $10\fdg5\times($cosh$((l-1\degr)/10\fdg5)-1)$ for the Northern bubble and $-8\fdg7\times($cosh$((l+1\fdg7)/8\fdg7)-1)$ for the Southern one. In \cite{Casandjian:2015p4414} we noticed that those catenary curves also reproduce correctly the structures observed close to the GC in the ROSAT X-ray observations \citep{BlandHawthorn:2003p4430}.

We also observe an extended excess of photons close to the GC at the base of the \fb. This feature is oriented nearly perpendicular to the Galactic plane and extend $\sim$4$\degr$ from the GC. In this energy range, several studies have reported emission from the GC region in excess of the expectations from standard emission components \citep{Vitale:2009p4234, Hooper:2011p4231, Abazajian:2012p4232, Calore:2015p4233}. We note that the shape of this excess is very uncertain for latitude less that 1\fdg5 due to systematics in the subtraction of foreground emissions.
%elated to the CR \isa{flux} and total hydrogen column densities close to the GC.

\section{The Galactic Interstellar Emission Model for the LAT}
\label{sec:Resulting_model_section}
To characterize point sources detected by the LAT, one requires a detailed spatial and spectral model of the total diffuse emission visible in their direction. We have used the component decomposition of Equation \ref{eqRing} and the results of its fits to the LAT data to build the GIEM that is publicly available at at the {\it Fermi} Science Support Center (FSSC) website\footnote{\url{http://fermi.gsfc.nasa.gov/ssc}}. The GIEM is meant to describe the total emission originating from the Galactic ISM, prior to its detection by the LAT, in other words, not convolved by the instrument response functions. The isotropic spectrum and the intensities due to the Sun, Moon, and Earth limb are not part of the GIEM, but distributed separately on the website because they depend on the photon selection and time span of observations.

%\subsection{A ne pas oublier de mettre quelque part}
%To construct the interstellar emission model we found templates for the gas column density and IC$_{p}$, we fitted Equation \ref{eqRing} to LAT photon count maps with  $q$, $C_{IC_{p}}$, $C_{iso}$, $C_{LoopI}$, $C_{patch_i}$, $C_{limb}$, $N_{pt_{i}}$ and $C_{ext_{i}}$ left free to vary in each energy bin and extrapolated the coefficients related to the hydrogen templates and $IC_{p}$ outside the energy range of the fit.

\subsection{Modeling the EEE}
\label{sec:modelization_of_the_large_scale_section}
In order to build a 3D intensity cube in position and energy, $I_{EEE}(l,b,E)$, of the EEE at angular scales relevant for point-source analyses, we have derived their spectral distributions in each sky pixel and we have used wavelet decomposition to retain structures on angular scales broader than $2\degr$. We have parametrized the spectral distributions by assuming IC interactions of a population of CR electrons with the CMB radiation. This parametrization is motivated by the presence of radio-emitting CR leptons in \loopI and the \fb \citep{Ade:2013p4413}, but its goal is to provide a flexible parametric form to describe the spectrum without Poisson noise in each pixel of the sky, rather than to interpret the spectra. We discuss the specifics of this procedure below.

To construct the $I_{EEE}$, we have rebinned the LAT count maps in each of the 14 energy bins to a $1\fdg8$ grid. In each pixel of this grid, we have fitted electron power-law spectra so that the sum of the baseline model (see section \ref{extra_emission}) and the supplementary electron IC emission reproduces the total photon count spectrum of the pixel. We need two independent power-law electron spectra in order to match the data over the whole $\gamma$-ray energy range from 50~MeV to 50~GeV. The power laws are respectively constrained by the LAT data below and above a $\gamma$-ray energy of 965~MeV. The resulting IC intensity maps are found in good agreement with those of the EEE presented in section \ref{extra_emission} between 50~MeV and 50~GeV. We have built skymaps of the power-law indices and normalizations for each electron population. To filter out point-like and small extended sources not present in our source list, as well as large Poisson fluctuations, we have transformed the spatial distributions of the electron spectral parameters into wavelets and filtered out scales smaller than 2$\degr$. Then applying the inverse transform, we have derived separate IC intensity maps from the low-energy and high-energy filtered electron spectra. We have applied a smooth spectral transition to merge the two IC distributions in each sky pixel. 
%We have not deconvolved by the LAT PSF the resulting IC map to ensure that no feature with an angular scale smaller than 2$\degr$ were incorporated into the model. We verified that this approximation does not bias the residual (en cas de questions aux referee).
Finally, in order to reduce the amount of LAT data to be reintroduced into the GIEM, we have restricted $I_{EEE}$ to regions where the EEE are bright. We have further verified at energies above 50~GeV, in five energy bins spanning from 50~GeV to 600~GeV, that the sum of $I_{EEE}$ and the baseline model agrees with the LAT observations. We note that we did not perform any deconvolution of the LAT counts map in order not to introduce structures with angular scales less than 2$\degr$ in the model.

%The physical motivation behind the production of those $\gamma$ rays in the energy range between 50~MeV and 50~GeV is not relevant if the resulting model is consistent with the data. For simplicity we favored a unique ISRF and $\gamma$-ray production process. 
%we have used the count excess maps obtained in each of the xxx energy bins above the model described at the beginning of section \ref{extra_emission}, which we momentarily refer to as the `base' model.  We have} deconvolved \isa{these maps} with the LAT PSF and rebinned them to a $1\fdg8$ grid. 
%used the count excess maps obtained in each of the xxx energy bins above the model described at the beginning of section \ref{extra_emission}, which we momentarily refer to as the `base' model.  We have} deconvolved \isa{these maps} with the LAT PSF and rebinned them to a $1\fdg8$ grid. \isa{
%The intensity we observe cannot be fitted by IC from an electron density represented by a simple power law over the whole $\gamma$-ray energy range from 50~MeV to 50~GeV. We then fitted two independent electron distributions, represented each one by a power-law spectrum, one corresponding to a low-energy fit below a $\gamma$-ray energy of 965~MeV and the other to a high-energy fit above this limit. 

In the right-hand column of Figure \ref{fig_comparaison_res_model} we show $I_{EEE}$ obtained at energies close to the geometric averages of the energy intervals use to display the count maps of the left column. We observe a good agreement between the large-scale structures in the LAT count maps and in the filtered component we have developed in order to account for these bright structures in the GIEM. In Figure \ref{gal_center}d we show a close-up view of the residuals integrated between 1.7~GeV and 50~GeV in the direction of the GC when $I_{EEE}$ is added to the baseline model. The residual map is flat apart from small-scale residuals toward the GC and the \fb. They correspond to the small angular scales not retained in the wavelet decomposition for the construction of $I_{EEE}$. 

\subsection{GIEM construction}
\label{sec:GIEM_construction_section}
We have built the GIEM by summing the emission components originating from the gas annuli and the DNM, allowing for the dust-related correction in \nhi, from the Galactic IC emission, and from $I_{EEE}$ (see Equation \ref{eqring3}). The gaseous components have been scaled according to the emissivity spectra $\frac{dq_{fit}}{dE}$ obtained in sections \ref{sec:local_annuli} and \ref{sec:inner_Gal_section} and parametrized with the pion decay and bremsstrahlung decomposition described in section \ref{sec:Gas_emissivities_section} to provide continuous functions. They are represented by solid lines in Figure \ref{gas_emiss}. The GALPROP IC distribution has been scaled with the normalization factors derived in section \ref{sec:ICnorm} and interpolated in energy as necessary (see Figure \ref{IC_renorm}).  We stress that %{\it Fermi}-LAT 
extended $\gamma$-ray sources with sizes larger than 2$\degr$ are partially incorporated into the GIEM by the addition of $I_{EEE}$, so they cannot be studied with this diffuse background model.
%We derived a final model for the interstellar emission from the sum of the modeled differential gas $\gamma$-ray emissivities ($\frac{dq_{fit}}{dE}$), the renormalized $I_{IC_{p}}$, and from the large-scale and extra emission $RES_{IC}$ (Equation \ref{eqring3}). 
%{\footnotesize
\begin{equation}
\begin{split}
  I(l,b,E) =  \sum_{i=HI,H_{2},DNM} \frac{dq_{fit_{i}}}{dE}(E) N_{H_i}(l,b)  +  {N}_{IC}(E)I_{IC_{p}}(l,b,E) + I_{EEE}(l,b,E)
\end{split}
\label{eqring3}
\end{equation} 
%}

\begin{figure}
\begin{center}
\includegraphics[width=12cm]{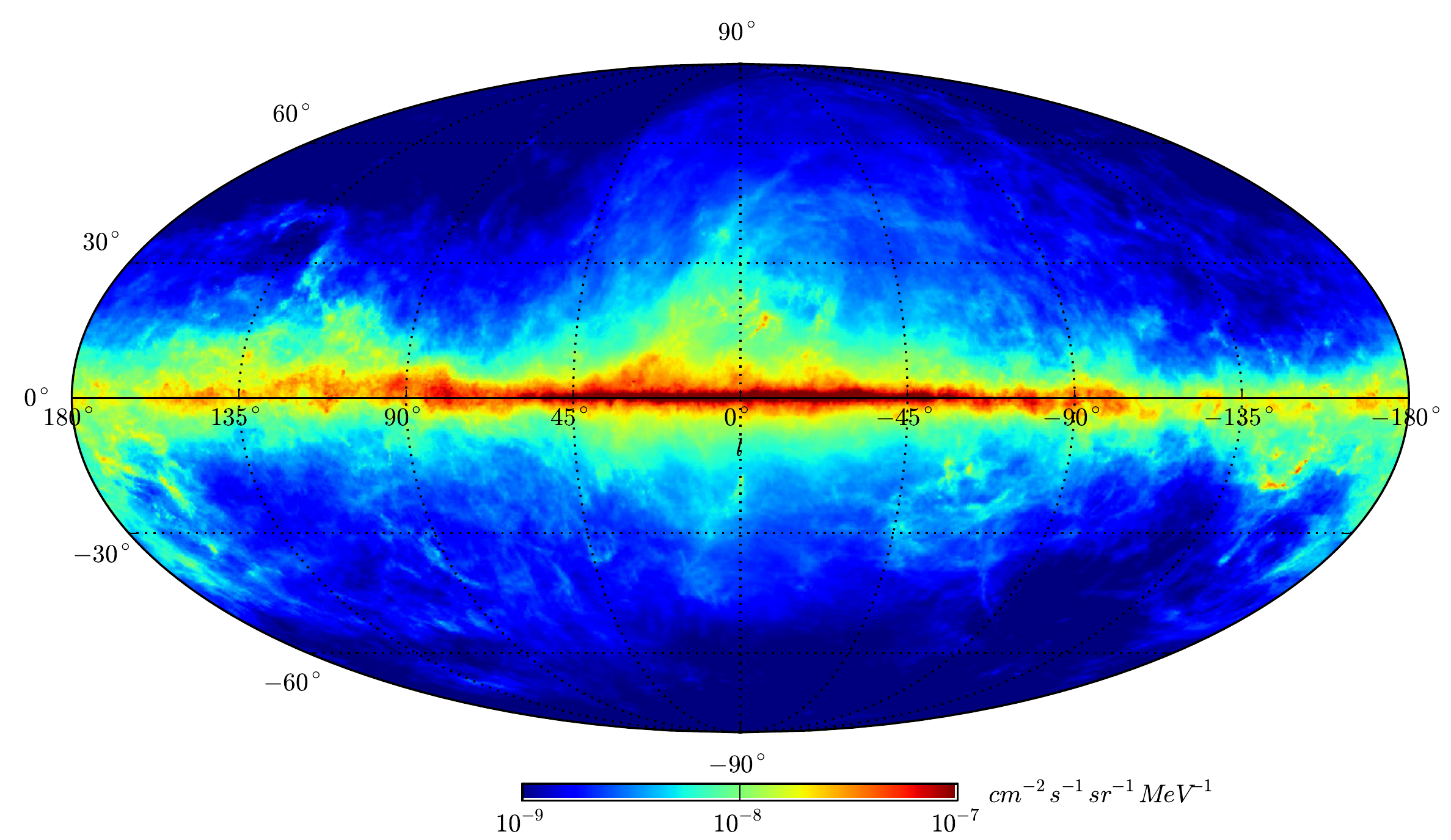}
\caption{Mollweide projection, displayed in log scaling, of the photon specific intensity in the Galactic Interstellar Emission Model at 1~GeV.}
\label{model_nosrc}
\end{center}
\end{figure}

The gas emissivities and IC normalization factors have been derived in the energy range 50~MeV to 50~GeV. Above this range, the low photon statistics do not allow for a reliable component separation, especially toward the Galactic plane. To build the GIEM for the range 50--600~GeV, we have extrapolated these factors as follows: 
\begin{itemize}
\item The spectral form chosen for the CR protons pervading the gas is equivalent at large rigidity to a simple power law of index $P_2$. We have checked the validity of this form to produce the right intensities beyond 50~GeV by comparing the LAT count maps recorded in 5 equal logarithmic bands between 50~GeV and 590~GeV to a model based on the extrapolation of the power-law proton spectra to several TV in rigidity. We have observed an excess of high-energy $\gamma$ rays in the LAT sky maps, including in the Galactic plane. To approximately account for this hardening, we have globally scaled the gas emissivity spectra above 50~GeV by the coefficient $0.8+(5.9\times10^{-3}\times E) - (2.8\times10^{-6}\times E^2)$ with the $\gamma$-ray energy $E$ in GeV.
This scaling would correspond to a gradual hardening for proton rigidities between approximately 200~GV and 2~TV. Near the Earth, the compilation of \textit{PAMELA} \citep{Adriani:2011p4018}, \textit{AMS-02} \citep{Aguilar:2015p4410}, \textit{CREAM} \citep{Yoon:2011p4411}, and \textit{ATIC-2} \citep{Panov:2006p4024} data indicate an up-turn in the proton spectrum starting around 500~GV. This break may signal a change in CR transport from diffusion on self-generated waves at low rigidities to diffusion on pre-existing magnetic turbulence at high rigidities \citep{Aloisio:2015p4412}.
The present need to scale up the local gas emissivity above 50 GeV to explain the LAT data at medium latitudes supports this CR hardening, but we defer a quantitative comparison to a dedicated study. The marked need for a larger high-energy intensity at low latitudes may be due to a comparable break in CR spectra and transport properties further out in the Galactic disc, but it can also stem from a population of unresolved hard sources, such as pulsar wind nebulae which are abundantly detected as TeV sources.

\item The Galactic IC intensity varies very smoothly across the sky. With increasing $\gamma$-ray energies, most of the IC emission occurs at low latitude toward the inner Galaxy. The lack of spatial variation at high latitude together with the low $\gamma$-ray statistics are such that the fit fails to reliably differentiate the IC from the isotropic emission above 50 GeV. 
Since the normalization of the GALPROP IC distribution is close to 1 at 50~GeV, we have relied on GALPROP predictions, without rescaling, for the extrapolation to higher energies.
\end{itemize}

\begin{figure}
\begin{center}
\includegraphics[width=12cm]{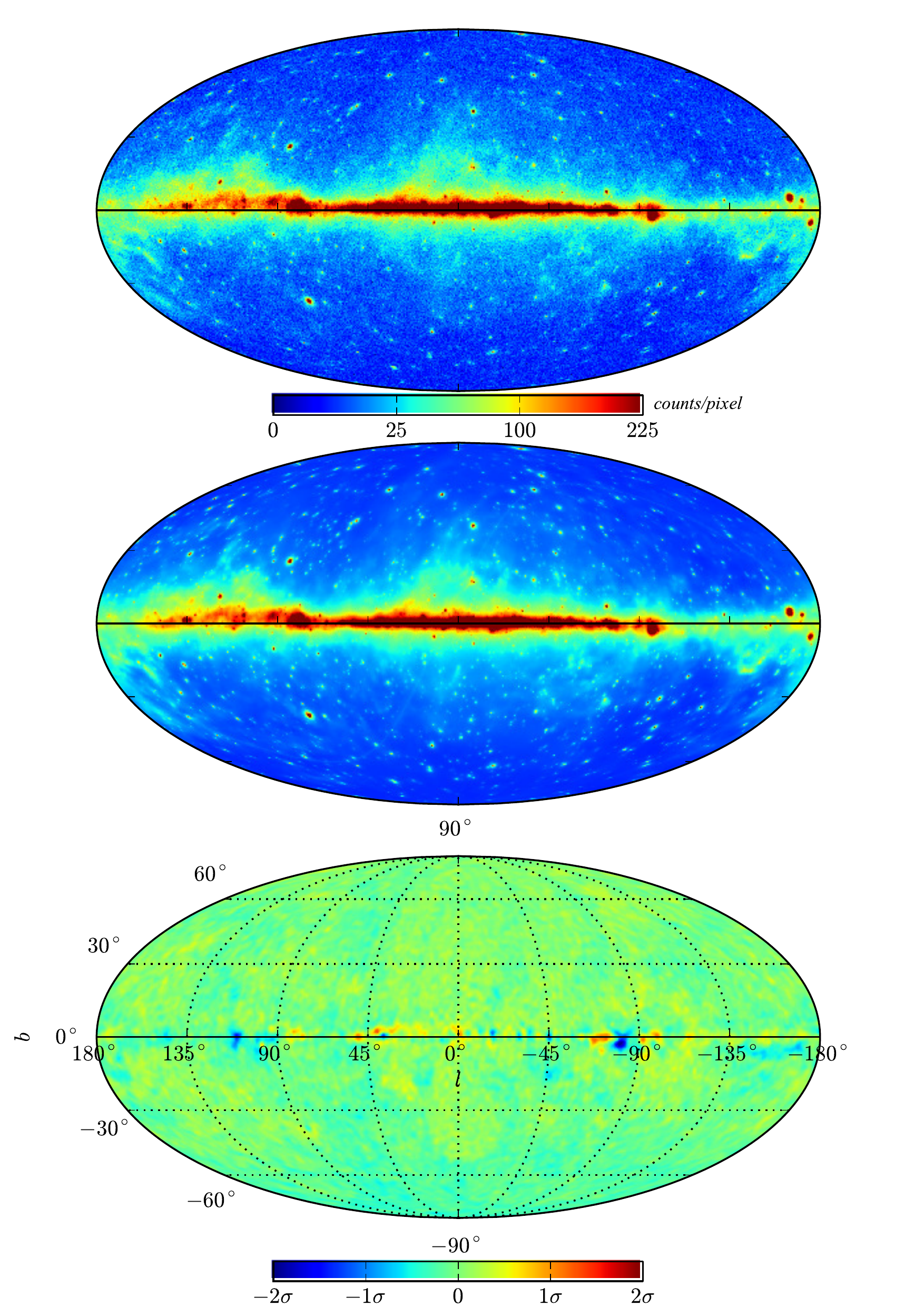}
\caption{Top: All-sky Mollweide projection for 4 years of {\it Fermi}-LAT $\gamma$-ray counts in the 0.36--50~GeV energy band. Middle: counts prediction in the same energy range based on the Galactic Interstellar Emission Model combined with modeled point and extended sources (including the Sun and the Moon), the residual Earth limb emission and the isotropic emission. Both maps are displayed with square root scaling to enhance emission away from the plane. Bottom: residual map in units of standard deviations after smoothing with a Gaussian of 2$\degr$ FWHM. The pixel size for the three maps is $0\fdg25$.}
\label{counts}
\end{center}
\end{figure}

We show in Figure \ref{model_nosrc} a map of the interstellar $\gamma$-ray emission in the GIEM at about 1~GeV.
In Figure \ref{counts} we compare the LAT count map integrated between 360~MeV and 50~GeV to the one predicted by the GIEM, combined with point and extended sources from a preliminary version of the 3FGL catalog, the isotropic emission and the emission from the Sun, the Moon, and from the residual Earth limb emission. 
%We derived the residual map (Figure \ref{counts}, bottom) by subtracting the model from the data and normalizing by the square root of the model to enhance deviations above statistical fluctuations. 
The overall agreement between observations and model is very good, partly because some of the detected excesses have been modeled and re-injected into the interstellar model. We still observe discrepancies along the Galactic plane at a level of less than 2$\sigma$.

%\begin{figure}
%\begin{center}
%\includegraphics[width=12cm]{residual_sigma.png}
%\caption{Residual map expressed in sigma values: $(N_{obs}-N_{pred})/ \sqrt{N_{pred}}$ }
%\label{residual}
%\end{center}
%\end{figure}

\begin{figure}
\begin{center}
\includegraphics[width=16cm]{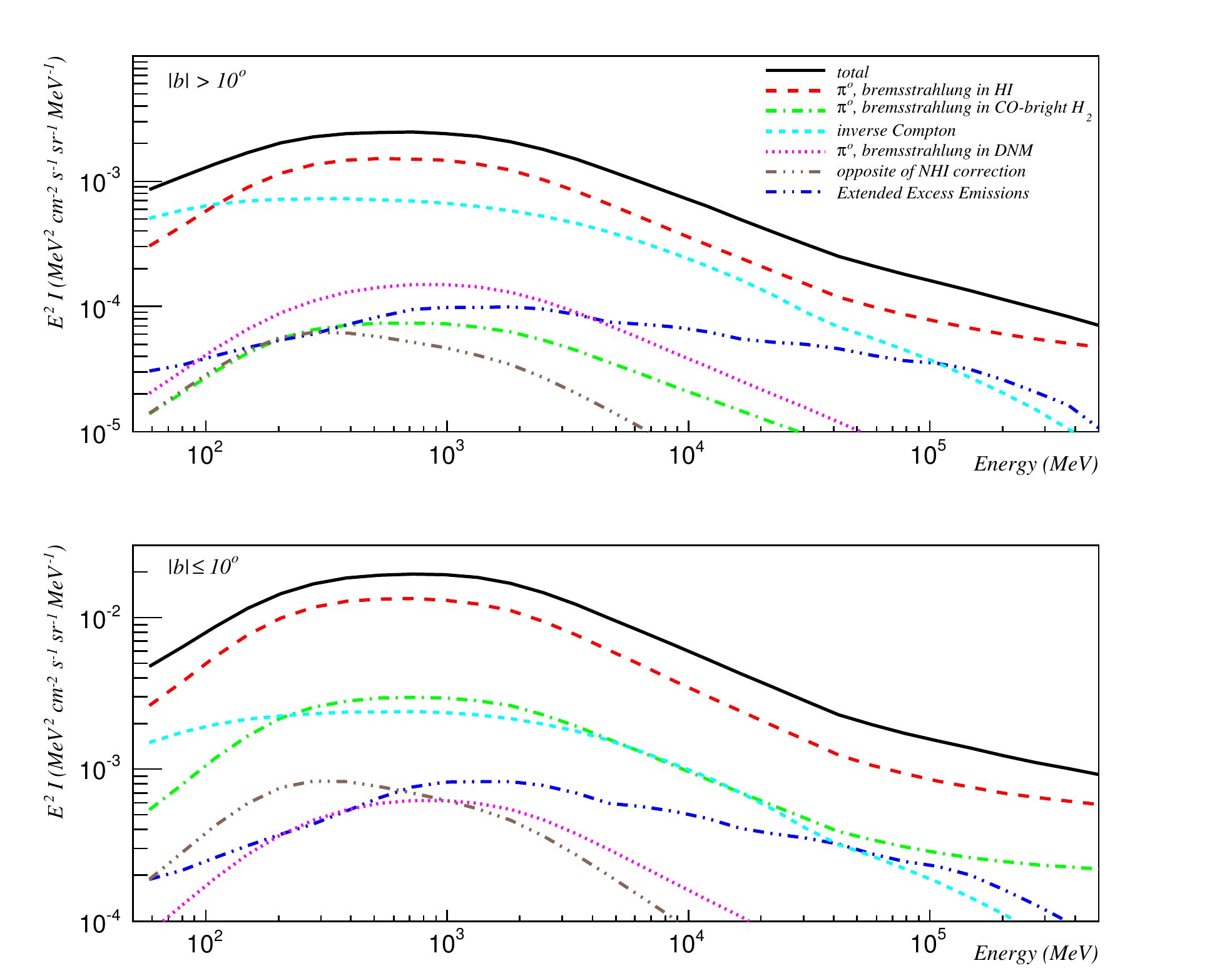}
\caption{Spectra of interstellar emission model components for $|b| >10\degr$ (upper panel) and $|b| < 10\degr$ (lower panel). We have decomposed the total intensity (solid line) into emission originating from hydrogen in its different phases: \hi (long-dashed), CO (dash-dotted), DNM (dotted). The emission from IC assuming an axisymmetric ISRF and electron distribution is shown as short dashed lines and the large-scale structures like \loopI and the \fb are shown as dashed-double-dotted. We show also the negative of the intensity associated with $N_{\text{H}\textsc{~i}}$ correction from the negative dust residual as dashed-triple-dotted.}
\label{model_I_spectre}
\end{center}
\end{figure}

In Figure \ref{model_I_spectre} we present the SEDs of various components of the GIEM, averaged over regions covering the Galactic disc ($|b| < 10\degr$) and higher latitudes. We have decomposed the total SEDs into contributions originating from the interstellar gas (atomic, molecular, and DNM), from an axisymmetric Galactic ISRF for the IC radiation, and from the EEE. We have represented the correction to the \hi contribution as an intensity by taking the negative of the photon intensity associated with the \nhi correction map. Above 100~MeV, %the hadronic decay following interactions of CR protons 
hadronic interactions with the atomic hydrogen dominate the interstellar $\gamma$-ray emission. The hardening of this emission above 50~GeV, obtained by scaling the gas emissivities as described above, is visible in both panels.  

The GIEM is available at the FSSC website as a FITS file named {\it gll\_iem\_v05\_rev1.fit}. For this file, we have resampled all the maps to a $0\fdg125$ grid in CAR projection, the format required by the LAT Science Tools analysis software. The FITS file comprises 30 logarithmically-spaced energies between 50 MeV and 600 GeV. It gives the photon specific intensity of the GIEM in photons sr$^{-1}$ s$^{-1}$ cm$^{-2}$ MeV$^{-1}$. This model tuned to LAT data is not corrected for the energy dispersion; therefore it can be used directly with LAT data. 
Version 15 (V15) of the P7REP IRFs is the recommended one for Pass 7 reprocessed data and for this model. The difference with the P7REP\_V10 set of IRFs used for the derivation of the gas emissivity spectra and IC normalization factors in this work mainly resides in an improved Monte Carlo PSF and in an updated fitting procedure to determine the parameters representing the LAT effective area. Those minor differences modify the exposure, but not the reconstructed LAT events. The minimum ratio in exposure (V15/V10) is 0.98 at 50~MeV and the maximum is 1.05 at 1~GeV. To correct the GIEM for the final (V15) IRFs, we have rescaled its intensity by the exposure ratio evaluated for each of the 30 energy planes of the GIEM. The model is then intended for use with the instrument response functions versions P7REP\_SOURCE\_V15, P7REP\_CLEAN\_V15, and P7REP\_ULTRACLEAN\_V15.

\subsection{GIEM accuracy}
The GIEM aims at a representation of the interstellar emission that closely reproduces the LAT observations. It combines three methods: the robust template-fitting method which uses no assumption on CR transport and spectra, but is sensitive to cross-correlations between the diffuse components and to source confusion at the lowest energies; the prediction of a propagation model (GALPROP) for the Galactic IC emission, with a scalable intensity in energy, but a fixed spatial distribution at each energy; and an iterative detection and data-based modelling of the EEE for which we have no external information. %\citep{Selig:2014p4318}. 
The interplay between those methods is such that deriving the uncertainties of our model is very challenging. 
%than for each individual method that results in a model less complete than the one describe in this work. The influence, on the inner annuli emissivities, of the iterative procedure described in Section \ref{sec:inner_Gal_section} is for example diffucult to quantify since we don't know what is missing in our model. Similarly the emissivities of the outer annuli depend on the spatial representation of \nhi that varies with the unknown distribution of $T_{S}$.

At high latitudes and outside the region covered by $I_{EEE}$, the uncertainties are likely dominated by the determination of the gas emissivities. The thinness of the local \hi is such that \nhi is not very sensitive to variations in $T_{S}$, so the uncertainties in the absolute determination of the LAT effective area dominate \citep{Casandjian:2015p4380}. 

Toward the outer Galaxy and outside the region covered by $I_{EEE}$, both uncertainties in the LAT effective area and in the uniformity of $T_{S}$ must be accounted for. We note that the column density in a line of sight can vary by up to a factor of 2 when assuming optically thin \hi or a $T_{S}$ of 95~K. The impact of a non-uniform $T_{S}$ is partially reduced by the \nhi correction applied with the dust-derived template. But the dust optical depths are easily biased by temperature confusion at low latitudes (see our discussion in section \ref{sec:Gas_emissivities_section}). The use of the `negative' \nhi correction map in the model improves the global fits to the LAT data, but can artificially lower the interstellar emission in a specific region. We recommend caution about potentially spurious features in the direction of hot dust or of steep gradients in dust temperature. 
%not totally as shown by the variation of the likelihood displayed in Figure \ref{TS}.  

%Inside the region shown in the right-hand column of Figure \ref{fig_comparaison_res_model}, 
In the other directions, the largest uncertainty in our model is its degree of incompleteness (large-scale sources off the plane like \loopI and the \fb, optically thick \hi and CO, poor gas distribution with distance, and missing DNM mass in the inner Galaxy). We have mitigated this incompleteness by including a component $I_{EEE}$ extracted from the data at angular scales broader than 2$\degr$.
%This was necessary since we have no accurate representation of \loopI, of the FBs or of the extra emission. Moreover, the dark gas is not included in the inner Galaxy region and $T_{S}$ is likely non-uniform with an average value potentially different from $T_{S}=140$~K measured toward the outer Galaxy.
But we observe in Figure \ref{counts} (bottom) that our model is still not complete since small deviations remain visible above the statistical fluctuations. This is a consequence of the interplay between the different components in the fits, which converge to the least-worse solution rather than the best one (a great improvement in one zone can be reached at the expense of minor worsening in others). It is also a consequence of adding a filtered map issued only from the positive residuals. We have also discussed in Section \ref{sec:Gas_emissivities_section} the uncertainties on the gas emissivities in the inner annuli. 

Although we cannot quantify the systematic uncertainties or degree of incompleteness of our model, we have assessed several indicators for the quality of this work:
\begin{itemize}
 \item coherent spectral distributions for the gas emissivities and IC normalization factors, in agreement with gas emissivity spectra previously obtained in dedicated studies of less confused regions, 
 \item coherent spatial structures of the EEE and $I_{EEE}$, strongly reminiscent of well-known features at other wavelengths, 
 \item a coherent and continuous shape for the edges of the \fb at low latitudes,
 \item the detection of the extended apparent path of the Sun and of the Moon across the sky in residual maps when they are not added to the model. 
 \item a flat final residual map within ${\pm}\,2\,\sigma$,
 \item the need for less than 5\% corrections to the GIEM specific intensities when fitting the data in the large majority of the 840 regions of interest used in the generation of the 3FGL catalog \citep[see Figure 25 of][]{Acero:2015p4385}.
 %For most of the 1728 ROIs used in the generation of the 3FGL, the diffuse emission normalization coefficients are within 10\% of their nominal values 
\end{itemize}

\section{Conclusion}
We have constructed a model for Galactic interstellar emission to allow the characterization of $\gamma$-ray point and small-extended sources in the LAT data with the best precision possible. The model is based on linear combinations of templates spatially correlated with production sites of $\gamma$ rays. We used \hi, CO, and dust reddening maps to describe $\gamma$ rays resulting from collisions between CRs and the ISM through hadrons decay and bremsstrahlung emission. The spatial distribution of $\gamma$ rays resulting from IC of CRs on the ISRF was calculated by the CR propagation code GALPROP. We determined the intensity associated with each template with a fit to LAT observations in several energy bands. In the first stage of the fit, extended emission like the \loopI and the \fb were accounted for by patches or through iterative procedures. This extended emission was included in the final model by re-injecting LAT residual counts above a baseline model. Those counts were filtered to only include structures with angular scales larger than 2$\degr$. The model is publicly available at the FSSC website.

We derived from this study the $\gamma$-ray emissivity spectrum at various Galactocentric distances. We interpreted those emissivities and observed that the spectrum of CR protons measured in the inner Galaxy is harder than in the outer Galaxy. We derived the radial distribution of the density of CR protons in the Galaxy, and find that it shows similarities with the distribution of tracers of massive star formation. In this work we characterized of the shape of the {\it Fermi} bubbles within 20$\degr$ from the plane and observed a non-centrosymmetric excess of $\gamma$ rays in the Galactic center above 1~GeV. We also observed a strong soft emission in the first and fourth quadrant from unknown origin.

% Acknowledgments
\acknowledgments
The \textit{Fermi} LAT Collaboration acknowledges generous ongoing support
from a number of agencies and institutes that have supported both the
development and the operation of the LAT as well as scientific data analysis.
These include the National Aeronautics and Space Administration and the
Department of Energy in the United States, the Commissariat \`a l'Energie Atomique
and the Centre National de la Recherche Scientifique / Institut National de Physique
Nucl\'eaire et de Physique des Particules in France, the Agenzia Spaziale Italiana
and the Istituto Nazionale di Fisica Nucleare in Italy, the Ministry of Education,
Culture, Sports, Science and Technology (MEXT), High Energy Accelerator Research
Organization (KEK) and Japan Aerospace Exploration Agency (JAXA) in Japan, and
the K.~A.~Wallenberg Foundation, the Swedish Research Council and the
Swedish National Space Board in Sweden.
 
Additional support for science analysis during the operations phase is gratefully acknowledged from the Istituto Nazionale di Astrofisica in Italy and the Centre National d'\'Etudes Spatiales in France.

Some of the results in this paper have been derived using the HEALPix \citep{Gorski:2005p1076} package.

% Bibliography
\bibliography{my_paper_template}
%\bibliography{paper_template}

\end{document}